\documentclass[a4paper,USenglish]{lipics-v2021}
\hideLIPIcs
\nolinenumbers
\usepackage[utf8]{inputenc}
\usepackage{amsmath, amssymb, amsthm}
\usepackage{booktabs}
\usepackage{enumerate}
\usepackage{hyperref}
\usepackage{enumitem}
\usepackage{xspace}
\usepackage[dvipsnames]{xcolor}
\usepackage{etoolbox}
\usepackage{tocloft}
\usepackage{listings}
\usepackage{cite}
\usepackage{microtype}
\usepackage{graphicx}

\lstdefinelanguage{Pseudo}{
  keywords={type, struct, array, size_t, bool, mutable, shared,
            function, return, if, else, foreach, private, process,
            void, NULL, true, false},
  keywordstyle=\bfseries,
  comment=[l]{//},
  commentstyle=\itshape\color{gray},
  morestring=[b]",
  sensitive=true
}
\makeatletter
\lst@AddToHook{OnEmptyLine}{%
  \advance\c@lstnumber-\@ne\relax%
}
\makeatother
\makeatletter
\renewenvironment{proof}[1][\proofname]{\par
  \pushQED{\qed}%
  \normalfont
  \topsep=1pt plus 0.5pt minus 0.5pt\relax  
  \trivlist
  \item[\hskip\labelsep\itshape #1\@addpunct{.}]\ignorespaces
}{%
  \popQED\endtrivlist\@endpefalse
}
\makeatother

\newcommand{\committee}{sending set\xspace}

\title{Decoupling Faults from Uncertainty: Message-Efficient Byzantine Agreement with Local Predictions}
\titlerunning{Decoupling Faults from Uncertainty}

\author{Naama Ben-David}{Technion -- Israel Institute of Technology, Haifa, Israel}{bendavidn@technion.ac.il}{}{}
\author{Tomer Shalit}{Technion -- Israel Institute of Technology, Haifa, Israel}{tomersha99@gmail.com}{}{}

\authorrunning{N. Ben-David and T. Shalit}
\Copyright{Naama Ben-David and Tomer Shalit}

\ccsdesc[500]{Theory of computation~Distributed algorithms}
\keywords{Byzantine agreement, algorithms with predictions, message complexity, distributed consensus, synchronous networks}



\begin{document}

\setlength{\abovedisplayskip}{3pt plus 1pt minus 1pt}
\setlength{\belowdisplayskip}{3pt plus 1pt minus 1pt}
\setlength{\abovedisplayshortskip}{1pt plus 1pt minus 1pt}
\setlength{\belowdisplayshortskip}{1pt plus 1pt minus 1pt}
\setlength{\textfloatsep}{7pt plus 2pt minus 2pt}
\setlength{\intextsep}{5pt plus 2pt minus 2pt}

\maketitle
\begin{abstract}
Byzantine Agreement protocols are traditionally designed under worst-case uncertainty about which processes are faulty, coupling the adversary's corruption budget $t$ with the system's epistemic uncertainty. Recent work showed that auxiliary failure predictions can improve round complexity, but cannot improve communication complexity when prediction errors are unconstrained. In this paper, we initiate a theoretical study of the complementary question: how much can Byzantine Agreement benefit when each process is guaranteed a local bound on its prediction error?

We study this question through the \emph{local $k$-error prediction model}, in which each process receives a local classification vector differing from the ground truth in at most $k$ positions. The predictions need not be common, and honest processes may hold substantially different views about whom to trust.

Under this promise model—where correctness relies on the local error bound holding—we present a synchronous Byzantine Agreement algorithm requiring $n \ge t + 4k + 1$, where $n$ is the total number of processes, $t$ is the maximum number of failures, and $k$ is the local error bound per honest process. In the regime where local uncertainty is small ($k \ll t$), our protocol decouples communication from the total fault budget $t$: in an execution where the actual maximum prediction error is $s \le k$ and the actual number of failures is $f \le t$, the protocol terminates in $O(s+1)$ rounds and sends $O((f+k)ks + nk)$ honest messages, achieving $o(n^2)$ communication when $t = \Theta(n)$ and $k = o(\sqrt{n})$. Finally, we show that it is not possible to solve Byzantine Agreement in the local $k$-error prediction model with fewer than $\Omega(nk)$ honest messages.
\end{abstract}

\newpage
\renewcommand{\figurename}{Algorithm}
\section{Introduction}\label{sec:intro}

Byzantine Agreement is a central problem in distributed computing, wherein processes in a distributed system must come to an agreement about a common output despite the presence of malicious processes that may act arbitrarily~\cite{pease1980reaching,lamport1982byzantine}. The Byzantine Agreement problem is traditionally studied under worst-case uncertainty about failures; a protocol is given only a bound $t$ on the number of Byzantine processes and must be correct no matter which processes are faulty. This classical framework inextricably couples two distinct dimensions:
the adversary's corruption budget $t$ (how many processes may fail) and the system's epistemic uncertainty (the ambiguity surrounding \emph{which} specific processes are faulty). In classical systems, uncertainty is maximal and equals $t$, causing fundamental communication lower bounds,
such as the $\Omega(nt)$ bound by Dolev and Reischuk~\cite{dolev1985bounds}, to scale directly with $t$.
Motivated conceptually by modern monitoring telemetry and anomaly-detection tools~\cite{darktrace,vectraai,zeek}, this raises a fundamental theoretical question: to what extent do the classical barriers of Byzantine Agreement stem from the number of faults $t$, versus the uncertainty about which processes are faulty?

Inspired by such prediction tools, as well as the growing body of work on algorithms with predictions~\cite{mitzenmacher22predictionscomm,predictionsList}, Ben-David et al. recently introduced the notion of Byzantine Agreement with predictions~\cite{ben2025byzantine}. In this model, processes are given predictions about the identity of faulty processes, and the protocol is required to remain safe even when the predictions are wrong. Ben-David et al.\ showed that predictions, even if they can be completely wrong, can improve round complexity when they are good, while follow-up work improved upon the bit complexity~\cite{dzulfikar2026communication} and fault tolerance~\cite{dallot2026resilient} of the original work. However, because these models permit unconstrained and arbitrary prediction errors, Ben-David et al.\ also showed an $\Omega(t^{2})$ message complexity lower bound for this setting~\cite{ben2025byzantine}, meaning that unbounded prediction models cannot break worst-case communication barriers when $t$ is large.

\subparagraph{Our Contributions.}
In this paper, we initiate a theoretical study of how \emph{bounded local uncertainty} about fault identities affects the resilience and communication complexity of Byzantine Agreement. Prior prediction-based BFT models follow the standard learning-augmented paradigm: correctness is preserved even when prediction errors are unbounded and arbitrary, at the cost of retaining worst-case communication barriers such as $\Omega(t^2)$. We study the complementary \emph{promise-model} regime, in which each process is guaranteed a local accuracy bound $k$ on its failure predictions. While practical behavioral detectors (such as intrusion detection or anomaly monitors~\cite{chandola2009anomaly,darktrace,kamvar2003eigentrust,vectraai,zeek}) generally cannot enforce a hard error bound against stealthy Byzantine adversaries that act honestly until a critical round, studying this idealized model cleanly isolates how epistemic uncertainty $k$ governs agreement complexity independently of the fault threshold $t$.

We formalize this question through the \emph{local $k$-error prediction model}. Each process receives a local classification vector predicting which processes are honest and which are Byzantine. These predictions need not be common: different honest processes may receive substantially different vectors and may therefore disagree about whom to trust. The only guarantee is local accuracy: each honest process’s prediction differs from the ground truth in at most $k$ positions. Thus, $k$ measures the amount of uncertainty in each honest process’s view of the failure pattern. This separates two quantities that are coupled in classical Byzantine Agreement: the number $t$ of possible Byzantine failures, and the amount $k$ of uncertainty about where those failures are. 

Our main result is a synchronous Byzantine Agreement protocol, titled \emph{Predictive Core-Leader Agreement (PCLA)}, whose performance
depends on the local prediction error. The protocol works whenever $n \ge t+4k+1$, with
no additional restriction on $t$. Unlike prior unbounded-error models, both the correctness and performance of PCLA rely on the local $k$-error guarantee holding (which correspondingly constrains an adaptive adversary's corruption choices, as formalized in Section~\ref{sec:model}). In particular, we focus on the fine-grained regime where local uncertainty is small ($k \ll t$), allowing the number of Byzantine processes $t$ to strictly surpass $n/3$. (When uncertainty is high, i.e., $k \ge \frac{1}{2}t$, our condition matches or exceeds $n > 3t$, where classical prediction-free protocols remain preferable.) In an execution where the actual maximum prediction
error of an honest process is $s\le k$, the protocol terminates in $O(s+1)$ rounds and sends
$O((f+k)ks+nk)$ honest messages, where $f\le t$ is the actual number of Byzantine processes.
In particular, when $t = \Theta(n)$ and $k = o(\sqrt n)$, PCLA achieves sub-quadratic $o(n^2)$ communication. Finally, we also show that it is not possible to solve Byzantine Agreement in the local $k$-error prediction model with fewer than $\Omega(nk)$ honest messages.

\begin{table}[t]
\centering
\small
\resizebox{\textwidth}{!}{
\begin{tabular}{lcccc}
\toprule
\textbf{Protocol} & \textbf{Prediction Model} & \textbf{Resilience ($n$)} & \textbf{Rounds} & \textbf{Honest Messages} \\
\midrule
Classical Early Stopping~\cite{berman1992optimal} & None & $n > 3t$ & $O(f+1)$ & $O(nt)$ \\
Ben-David et al.~\cite{ben2025byzantine} & Global error $B$ & $n > 3t$ & $O(\min\{B/n, f\} + 1)$ & --- \\
Dzulfikar \& Gilbert~\cite{dzulfikar2026communication} & Global error $B$ & $n > 3t$ & $O(\min\{B/n, f\} + 1)$ & $O(n^2 \log(\min\{B/n, f\} + 1))$ \\
\textbf{This Work (PCLA)} & Local error $k$ & $\mathbf{n \ge t + 4k + 1}$ & $\mathbf{O(s+1)}$ & $\mathbf{O((f+k)ks + nk)}$ \\
\bottomrule
\end{tabular}
}
\caption{Comparison of synchronous unauthenticated Byzantine Agreement protocols. Here, $t$ is the fault threshold, $f \le t$ is the actual number of failures, $B$ is the aggregate prediction error across honest nodes, $k$ is the local error bound per node, and $s \le k$ is the actual maximum local error in an execution.}
\label{tab:comparison}
\end{table}

The protocol uses the local prediction guarantee structurally. Each honest process orders the processes according to its own prediction, and communicates only with a small predicted-honest prefix. Although these prefixes may differ, the local $k$-error condition ensures that they have large overlap with the true honest prefix, and with each other. This is the basis for all communication in the protocol; processes that place themselves outside of the honest prefix do not participate, while the rest participate in a leader-driven algorithm to come to an agreement, drawing conceptual inspiration from the Phase King paradigm ~\cite{berman1989towards,berman1992optimal}. The eventual existence of a common honest leader is guaranteed by the overlap of the prefixes of different processes. 

\section{Related Work}\label{sec:related}

Byzantine Agreement was introduced by Lamport, Shostak, and Pease~\cite{pease1980reaching,lamport1982byzantine}. In the synchronous unauthenticated setting, deterministic agreement is possible iff $n>3t$, while authenticated settings admit higher resilience~\cite{dolev1983authenticated}. In the asynchronous case, authentication does not help, but trusted hardware can increase fault tolerance~\cite{aguilera2019impact,aguilera2023ubft,aguilera_et_al:LIPIcs.DISC.2021.3,ben2022revisiting,clement2012limited,chun2007attested,zeno2026link}.

Our protocol is most closely related to phase-king algorithms~\cite{berman1989towards,berman1992optimal,lenzen2022recursive}. As in the phase-king paradigm, agreement is driven by the eventual emergence of an honest leader. Unlike classical protocols, however, candidate kings are derived from local predictions rather than a globally fixed sequence.
A second key ingredient is graded agreement~\cite{feldman1988optimal}, which provides a similar abstraction to adopt-commit~\cite{gafni1998round} and crusader agreement~\cite{dolev1982byzantine}. All of these primitives relax agreement to allow processes to express a degree of confidence in their decision value.
Graded agreement and its variants have been widely studied, improving efficiency and strengthening guarantees in a variety of models~\cite{abraham2022efficient,abraham2023round,civit2024efficient,aspnes2014tight}. We use a graded-agreement subroutine adapted to prediction-induced communication neighborhoods.

Dolev and Reischuk proved fundamental communication lower bounds for Byzantine Agreement, including an $\Omega(nt)$ bound in the unauthenticated setting~\cite{dolev1985bounds}. In our model, the analogous parameter is the prediction error $k$: we prove an $\Omega(nk)$ lower bound.

\medskip
\noindent
\textbf{Algorithms with Predictions.} Our work fits within the framework of algorithms with predictions, or learning-augmented algorithms~\cite{mitzenmacher22predictionscomm,predictionsList}. In this paradigm, algorithms exploit possibly erroneous predictions while retaining worst-case guarantees. The framework has been applied to many online and approximation problems, including caching, ski rental, scheduling, matching~\cite{ahmadian23loadbalancing,ANGELOPOULOS24onlineuntrustedadvice,mitzenmacher22predictionscomm,purohit2018improving,azar2022online}, and recently also distributed graph algorithms~\cite{boyar2025predictions}.

Most works in this paradigm assume the predictions can be arbitrarily unreliable. However, a few recent works have suggested attaching metrics to the uncertainty one has about the predictions~\cite{shen2025algorithms,sun2024online}, in a somewhat similar vein to our model.

\medskip
\noindent
\textbf{Byzantine Agreement with Predictions.}
The closest prior work is Byzantine Agreement with predictions, introduced by Ben-David, Dzulfikar, Ellen, and Gilbert~\cite{ben2025byzantine}. 
Subsequent work improves communication complexity~\cite{dzulfikar2026communication} and studies resilience-performance tradeoffs under predictions~\cite{dallot2026resilient}. In particular, Dallot et al.~\cite{dallot2026resilient} established consistency-robustness trade-offs for global predictions and showed that under arbitrary local predictions, achieving resilience beyond $n/2$ is impossible. These works follow the standard learning-augmented approach: correctness is independent of prediction quality, while performance improves as predictions become more accurate.

Our work studies a complementary notion of prediction quality. Rather than bounding aggregate error while requiring robustness to arbitrary predictions, we study a promise model in which each honest process's prediction differs from the ground truth in at most $k$ positions and protocol correctness depends on this guarantee. This stronger assumption allows predictions to be used structurally to circumvent the $\Omega(t^2)$ communication barrier and the $n/2$ local resilience impossibility of the unbounded-error setting.

\medskip\noindent
\textbf{Failure Detectors.}
The idea of augmenting agreement algorithms with information about failures in the system dates back to the 90s, with the introduction of unreliable failure detectors by Chandra and Toueg~\cite{chandra1996unreliable,chandra1996weakest}. Failure detectors typically provide evolving suspicions about crash failures and are characterized by completeness and accuracy properties. In contrast, our model provides static predictions about Byzantine identities, with quality measured by local Hamming distance from the ground truth. We are interested not in overcoming asynchrony, but in understanding how uncertainty about Byzantine identities affects agreement complexity.

Related work also studies Byzantine failure detection, where Byzantine processes can eventually be identified or isolated~\cite{malkhi1997unreliable}. Our setting differs in that predictions are prior information, may be heterogeneous, and can contain both false positives and false negatives.

\medskip\noindent
\textbf{Asymmetric Trust and Heterogeneous Quorum Systems}.
Another related direction studies consensus under asymmetric trust assumptions. Federated Byzantine Agreement, introduced by Stellar~\cite{mazieres2015stellar}, allows different nodes to trust different quorum slices, and safety depends on the resulting quorum-intersection structure. Subsequent work develops the theory of asymmetric Byzantine quorum systems and asymmetric consensus~\cite{cachin2021asymmetric}.

Like these models, our setting gives different honest processes different local trust views. However, the asymmetry arises from potentially incorrect predictions rather than explicit trust assumptions. The required intersection properties are therefore derived from the local $k$-error guarantee rather than assumed as part of the system specification.
\section{Model}\label{sec:model}

\textbf{Processes.} We consider a system with $n$ processes $\{p_1,\dots,p_n\}$. Up to $t$ of these processes may be faulty, or \emph{Byzantine}, meaning that they can act arbitrarily. The remaining processes are \emph{honest}, meaning that they always follow the protocol. For a given execution, let $f\le t$ be the actual number of faults. (The parameter $f$ is not known to the protocol while $t$ and $n$ are.) We use $\mathcal{H}$ to denote the set of honest processes' identifiers, and $F=\{1,\dots,n\}\setminus \mathcal{H}$ as the set of faulty processes' identifiers. We consider a constrained adaptive adversary that may corrupt up to $t$ processes during the execution, subject to the local $k$-error constraint defined below.

\textbf{Network Assumption.} Communication is synchronous and all-to-all: any message sent by process $p$ to $q$ in round $r$ arrives within round $r$ (within bounded delay $\Delta$).

\begin{definition}[Byzantine Agreement]
Each process $p_i$ begins with an input value $v_i$ and must produce a decision satisfying:
(1)~\textbf{Agreement}: all honest processes decide the same value;
(2)~\textbf{Validity}: if all honest processes start with $v$, all decide $v$; and
(3)~\textbf{Termination}: all honest processes eventually decide.
\end{definition}

Round complexity measures the rounds until the last honest process decides; message complexity measures the total messages sent by honest processes. Both are evaluated in the worst case over all inputs and adversarial strategies.

\medskip\noindent
\textbf{Predictions.} In addition to standard inputs, each process $p_i$ receives a local prediction string $a_i \in \{0, 1\}^n$ prior to execution. These predictions may be heterogeneous ($a_i \neq a_\ell$). We focus on \textit{classification predictions}~\cite{ben2025byzantine}, where $a_i[j]=1$ predicts $p_j$ is honest and $a_i[j]=0$ predicts $p_j$ is faulty. We define the ground truth vector $\hat{a} \in \{0, 1\}^n$ such that $\hat{a}[j]=1$ if and only if $p_j$ is honest. A prediction bit $a_i[j]$ is considered correct if $a_i[j] = \hat{a}[j]$. Additionally, we assume that every honest process always identifies itself as honest.

\medskip\noindent
\textbf{Local $k$-Error Constraint.}
For every honest process $p_i \in \mathcal{H}$, the Hamming distance between its initial prediction vector $a_i$ and the ground truth $\hat{a}$ (defined with respect to the final faulty set $F$) is at most $k$. This error budget $k$ may be comprised of False Positives (FP) or False Negatives (FN). Accordingly, the adversary may corrupt processes adaptively during the execution as long as $|F| \le t$ and this $k$-error bound holds for every $p_i \in \mathcal{H}$—constraining the admissible corruption sets while preserving full adaptive control over corruption timing and Byzantine behavior.

\medskip\noindent
\textbf{Ordering notation.}
Let $\pi: \{0,1\}^n \to S_n$ denote the mapping from a classification vector to a permutation of the process identifiers $\{1, \dots, n\}$. In this ordering, processes classified as honest ($a[j]=1$) precede those classified as faulty ($a[j]=0$), with both groups maintaining their relative increasing order by process ID. Formally, for any classification vector $a$, the position of process $p_j$ in the permutation $\pi(a)$ is:

\begin{equation}
\pi(a)(j) = 
\begin{cases} 
\sum_{r=1}^j a[r] & \text{if } a[j]=1 \\[8pt]
j + \sum_{r=j+1}^n a[r] & \text{if } a[j]=0 
\end{cases}
\end{equation}

We define $\hat\pi = \pi(\hat{a})$ as the ground-truth permutation, and for each process $p_i$, we denote its local prediction permutation as $\pi_i = \pi(a_i)$.

\medskip\noindent
\textbf{Prefix Sets.} For every process $p_i$ and integer $x$, let
$\Pi_i^x=\{p_j : \pi_i(j)\le x\}$
denote the first $x$ processes in the local ordering of $p_i$. Similarly, let
$\hat\Pi^x=\{p_j : \hat\pi(j)\le x\}$
denote the first $x$ processes in the ground-truth ordering.

\begin{observation}\label{obs:pihat correct}
    In a system with $n \ge x + t$, for any $y \le x$, all processes in $\hat\Pi^y$ are honest.
\end{observation}
\section{Algorithmic Overview and Setup}
\label{sec:overview}
At its core, the local $k$-error prediction BFT algorithm builds upon the leader-driven coordination paradigm of the Phase King algorithm~\cite{berman1989towards}. That is, in each round, some candidate leader is chosen among the processes, and that leader attempts to make everyone reach a decision. A graded agreement algorithm is executed twice in each round, to allow for a safe early-stopping mechanism.

However, unlike classic Phase King, the candidate leader is not chosen in a round robin fashion. Instead, each process $p_i$ uses its prediction to create the local ordering $\pi_i$, in which honest-predicted processes are placed before those that are predicted to be faulty. Intuitively, it uses a prefix of this ordering, namely, $\Pi_i^{4k+1}$, as its candidates for leader, trying a new candidate in each round of the consensus algorithm. To decrease the number of inconsistencies in the candidates of different processes, the prefixes are shared in a dissemination phase, similar to what is done in the algorithm of Ben-David et al~\cite{ben2025byzantine}. However, our algorithm decreases message complexity by having each process limit its communication to a subset of the other processes, which we show, using the local $k$-error guarantee, has sufficient overlap with other processes' communication subsets to ensure quorum intersection.

\subsection{Listening and Sending Sets}\label{sec:send-listen}
To reduce the communication, our algorithm relies on a few key ideas. Firstly, due to the guarantees on the quality of the received predictions, we are able to limit the number of honest processes that participate in the protocol. 
Specifically, we define the \emph{$x$-core set} $C_x = \{p_i : \pi_i(i)\le x+k\}$, containing processes that place themselves in the first $x+k$ positions of their local ordering.

Notably, each process locally knows whether it belongs to $C_x$ without having to communicate with any other process.
The protocol then makes use of the $4k+1$-core for all of its communication. 
Any honest process outside of this core set does not participate in the protocol, and instead simply waits to be notified of the decision. 
Since $C_{4k+1}$ is used throughout the protocol, we simply refer to it as the \emph{core set} of processes. Furthermore, honest processes in the core set, also referred to as \emph{honest core processes}, only ever listen to messages received from the first $4k+1$ processes in their local ordering (denoted $\Pi_i^{4k+1}$ for process $p_i$). Almost all other messages they receive are ignored, with a few notable exceptions. We therefore sometimes refer to $\Pi_i^{4k+1}$ as $p_i$'s \emph{listening set}. Due to lack of space, in Appendix~\ref{app:overlap}, we show the following key properties about the overlap of these sets with each other and with the set of honest processes.

\begin{lemma}
\label{lem:pi hat intersect}
For any honest process $p_i$ and any $x\le n-t$,
\[
|\Pi_i^x \cap \hat\Pi^x| \ge x-k.
\]
\end{lemma}
Intuitively, this lemma holds due to the local $k$-error prediction property; the orderings inherit the $k$ inconsistency from the Hamming distance of the prediction vectors. To formally show this somewhat intuitive yet non-trivial property, we account for the effect of false negative and false positive errors in the prediction string of $p_i$ separately. From there, arguing about set overlap leads to the following key lemma.

\begin{lemma}[Honest core size]\label{lem:C size}
For any $x\le n-t$, the set $C_x$ contains all the first $x$ processes in $\hat\pi$, i.e.\ $\hat\Pi^x\subseteq C_x$, and 
$|C_x\cap \mathcal H|\le x+2k$.
\end{lemma}

The following corollary shows that there is significant overlap in the listening sets of honest processes, and in particular, that there is a quorum of size at least $2k+1$ of honest processes of overlap between any two honest processes' listening sets.

\begin{corollary}[Listening-set overlap]
\label{cor:listening-overlap}
For any two honest processes \(p_i,p_j\),
\(|\Pi_i^{4k+1}\cap \Pi_j^{4k+1}\cap\mathcal H|\ge 2k+1\).
\end{corollary}

Note that while each process knows whether it is part of the core set, and which processes it should listen to when receiving messages, it \emph{does not} know who else is in the core set, processes can only check their own membership. Processes therefore have to determine who to send their messages to, with the goal being that each process $p_i$ send its messages to exactly the honest core processes that will listen. For this purpose, the protocol begins with a set up phase in which each process $p_i$ in the honest core sends a `ping' to each of the first $4k+1$ processes in its local order, denoted $\Pi_i^{4k+1}$. Each honest core process $p_i$ then collects all received pings into its \emph{\committee} $Send_i$, and later sends all its messages to all processes in $Send_i$. Note that $Send_i$ will include all honest core processes $p_j$ such that $p_i\in \Pi_j^{4k+1}$, but may also include up to $t$ Byzantine processes. Algorithm~\ref{alg:init} presents this set up phase.

\begin{figure}[t]
\begin{minipage}[t]{0.47\textwidth}
\begin{lstlisting}
Code for process @$p_i$@:
    @$Send_i = \emptyset$@
initialize():
    if @$p_i \in C_{4k+1}$@: //equivalently, if @$\pi_i$@(i) @$\leq$@ 5k+1
        send ping to all in @$\Pi_i^{4k+1}$@
        upon receiving ping from process @$p_j$@:
            add @$p_j$@ to @$Send_i$@
\end{lstlisting}
    \caption{Set Up Protocol -- building the \committee}
    \label{alg:init}
\end{minipage}
\hfill
\begin{minipage}[t]{0.50\textwidth}
\begin{lstlisting}
@$cc_i$@ //n-bit array
generateCandidates():
    receivedPreds = @$\emptyset$@
    if @$p_i\in C_{4k+1}$@:
        send @$\Pi_i^{4k+1}$@ to all in @$Send_i$@
    receivedPreds = msgs received from @$\Pi_i^{4k+1}$@
    for each j between 1 and n:
        if @$\ge 2k+1$@ msgs predict @$p_j$@ honest:
            @$cc_i$@[j] = 1
        else:
            @$cc_i$@[j] = 0
\end{lstlisting}
    \caption{Candidate Leader Generation}
    \label{alg:classification}
\end{minipage}
\end{figure}

The following lemma formally proves the intuitive statement above: the \committee{s} are formed correctly, such that all honest processes in an honest process $p_i$'s listening set place $p_i$ in their sending set.

\begin{lemma}[Sending-Listening Overlap]
\label{lem:overlap}
Any honest process $p_i\in C_{4k+1}$ is included in the \committee{s} of at least $3k+1$ honest processes 
from $\hat\Pi^{4k+1}$.
\end{lemma}
\begin{proof}
From the protocol, every process $p_i\in C_{4k+1}$ will send ping to its first $4k+1$ processes ($\Pi_i^{4k+1}$). By Lemma~\ref{lem:pi hat intersect}, at least $3k+1$ of them overlap with $\hat\Pi^{4k+1}$. By Observation~\ref{obs:pihat correct}, all those in the overlap are honest. 
Furthermore, note that by Lemma~\ref{lem:C size}, all those in the overlap are in $C_{4k+1}$ so they will place $p_i$ in their \committee{s}.
\end{proof}

\subsection{Candidate Leader Generation}\label{sec:classification}

The next key element of the protocol involves the creation of a list of potential candidate processes that can serve as leader in PCLA.

Intuitively, the ideal candidate set is $\hat\Pi^{4k+1}$, the first $4k+1$ processes in the ground-truth ordering. However, each process $p_i$ only has access to its local ordering $\Pi_i$. Ideally, process $p_i$ would use its listening set $\Pi_i^{4k+1}$ as a potential candidate set to approximate $\hat\Pi^{4k+1}$. While the listening sets of honest processes have large overlap (by Corollary~\ref{cor:listening-overlap}), this overlap is not enough to guarantee that processes will align on the same honest leader within a limited number of rounds.

Thus, to improve this overlap and guarantee that there will be some honest process trusted by all honest core members, honest core processes report their listening sets to each other through their \committee{s} and employ a threshold rule to update their trust in each process in the set and generate a candidate list for PCLA. More specifically, each honest core process $p_i$ builds a \emph{classification string} $cc_i$, where $cc_i[j] = 1$ if $p_i$ considers $p_j$ to be a candidate leader, and $cc_i[j] = 0$ otherwise. To do so, each core process sends its listening set to the processes in its sending set $Send_i$, and each process classifies $p_j$ positively (i.e., adds it to its candidate set) if $p_j$ appears in at least $2k+1$ of the received listening sets. 
This threshold has two roles. First, it guarantees that some honest process in $\hat\Pi^{4k+1}$ is
classified positively by every honest process. Second, it prevents too many processes
outside $\hat\Pi^{4k+1}$ from being classified positively. These two properties allow PCLA to search over a bounded set of candidate leaders
while ensuring that the search eventually reaches a common honest candidate.
This phase is similar to the classification phase used in Ben-David et al.~\cite{ben2025byzantine}. The pseudocode for the candidate generation phase is presented in Algorithm~\ref{alg:classification}.

We now show the useful properties discussed above, starting with showing that there is a common honest candidate for all honest processes. Its full proof appears in Appendix~\ref{app:candidates}.

\begin{lemma}[Common honest candidate]
\label{lem:common honest}
There exists an honest process $p^\star\in \hat\Pi^{4k+1}$ such that every honest core process classifies it positively:
$\forall p_i\in\mathcal H \cap C_{4k+1}:cc_i[p^\star]=1$.
\end{lemma}
\begin{proof}[Proof sketch]
Every process in \(\hat\Pi^{4k+1}\) is honest and belongs to the core. Moreover,
by Lemma~\ref{lem:pi hat intersect}, each such process \(p_r\) has at least
\(3k+1\) members of \(\hat\Pi^{4k+1}\) in its listening set. Thus the processes
of \(\hat\Pi^{4k+1}\) cast many votes for other processes in
\(\hat\Pi^{4k+1}\): in total, at least \((4k+1)(3k+1)\) such votes.
Since these votes are distributed among only \(4k+1\) processes, some
\(p^\star\in\hat\Pi^{4k+1}\) receives at least \(3k+1\) votes from processes in
\(\hat\Pi^{4k+1}\). Now consider any honest core process \(p_i\). Its listening
set also contains at least \(3k+1\) processes from \(\hat\Pi^{4k+1}\). Therefore
\(p_i\)'s honest-prefix listeners intersect the \(3k+1\) voters for \(p^\star\)
in at least \(2k+1\) processes. Hence \(p_i\) receives at least \(2k+1\) votes
for \(p^\star\), and so \(cc_i[p^\star]=1\). Since \(p_i\) was arbitrary, every
honest core process classifies \(p^\star\) positively.
\end{proof}

The next lemma quantifies how much disagreement the classification phase can create.
There are two possible types of classification error relevant to the candidate leader
argument. First, a true member of $\hat\Pi^{4k+1}$ may fail to appear in every
honest classification vector; these are the missing true candidates. Second, a
process outside $\hat\Pi^{4k+1}$ may appear in some honest classification vector;
these are the extra candidates.

The key point is that both types of errors must be paid for by bad prefix incidences
among the processes in $\hat\Pi^{4k+1}$. A missing true candidate requires many
honest-prefix senders to omit it, while an extra candidate requires many
honest-prefix senders to include it. Thus the total number of problematic
classification outcomes is controlled by the total prediction error inside $\hat\Pi^{4k+1}$.
We define the classified false negatives by $\mathrm{FalseNeg}_{4k+1} = \hat\Pi^{4k+1} \cap \left(\bigcup_{p_i\in C_{4k+1}\cap\mathcal H} \{p_x : cc_i[x]=0\}\right)$, and the classified false positives by $\mathrm{FalsePos}_{4k+1} = \left(\bigcup_{p_i\in C_{4k+1}\cap\mathcal H} \{p_x : cc_i[x]=1\}\right) \setminus \hat\Pi^{4k+1}$.
Thus $\mathrm{FalseNeg}_{4k+1}$ contains the true-prefix processes rejected by
at least one honest core process, while $\mathrm{FalsePos}_{4k+1}$ contains the
outside-prefix processes accepted by at least one honest core process.
The following lemma bounds the total classification error, and its proof appears in Appendix~\ref{app:candidates}.
Recall that \(s\le k\) denotes the actual maximum prediction error of an honest
process in the execution.

\begin{lemma}[Global classification error bound]
\label{lem:classification-error-budget}
$|\mathrm{FalseNeg}_{4k+1}|+|\mathrm{FalsePos}_{4k+1}|\le 8s$.
\end{lemma}

\begin{proof}[Proof sketch]
We charge each classification error to prediction mistakes made by processes in
the true prefix \(\hat\Pi^{4k+1}\). If a true-prefix process \(p_j\) is rejected
by some honest core process \(p_i\), then \(p_i\) saw fewer than \(2k+1\) votes
for \(p_j\). Since \(p_i\)'s listening set contains at least \(3k+1\) processes
from \(\hat\Pi^{4k+1}\), at least \(k+1\) true-prefix senders must have omitted
\(p_j\).

Similarly, if a process \(p_j\notin\hat\Pi^{4k+1}\) is accepted by some honest
core process $p_i$, then \(p_i\) saw at least \(2k+1\) votes for \(p_j\). Since at
most \(k\) of these votes can come from outside \(\hat\Pi^{4k+1}\), at least
\(k+1\) true-prefix senders must have incorrectly included \(p_j\).

Thus each false negative or false positive consumes at least \(k+1\) bad prefix
incidences. Each process in \(\hat\Pi^{4k+1}\) has at most \(s\) prediction
errors, so its local prefix can differ from \(\hat\Pi^{4k+1}\) in at most \(s\)
extra and \(s\) missing processes. Hence the total number of bad prefix
incidences is at most \((4k+1)\cdot 2s\), and the number of classified errors is
at most $\frac{(4k+1)2s}{k+1}\le 8s$.
\end{proof}

\begin{remark}[Prefix-sensitive refinement]\label{rem:sigma-defintion}
The charging argument only uses prediction mistakes affecting coordinates up to the largest identifier $x$ in $\hat\Pi^{4k+1}$.  For a prediction vector \(a\), let \(a|_{[x]}\) denote its restriction to the coordinates \(\{1,\ldots,x\}\), and let \(d_H\) denote Hamming distance. Defining the average actual prefix error as $\sigma = \frac{1}{4k+1}\sum_{p_r\in\hat\Pi^{4k+1}} d_H(a_r|_{[x]},\hat a|_{[x]})$, the same bound yields $|\mathrm{FalseNeg}_{4k+1}|+|\mathrm{FalsePos}_{4k+1}| = O(\sigma)$. This prefix-sensitive quantity can be smaller than $s$, but we state main bounds in terms of $s$ for clarity.
\end{remark}
\section{The Predictive Core-Leader Agreement (PCLA) Protocol}\label{sec:phaseking}

We can now present our protocol, Predictive Core-Leader Agreement (PCLA), which builds upon the structure of leader driven consensus protocols~\cite{berman1989towards,berman1992optimal} while leveraging local $k$-error-bounded predictions to improve round and message complexity. At a high level, honest core processes repeatedly send their candidate set to all processes in their \committee, wait to receive candidate sets from all processes in their listening set, and use these candidate sets to determine (a) whether to act as a leader in this round, and (b) whether to accept a leader proposal from any given process. The candidate sets are then updated at the end of the round, eliminating all processes that lack sufficient support, or have been proven Byzantine. 

\subsection{Graded Agreement Black Box}\label{sec:graded}
The protocol uses a graded agreement protocol as a building block. Our graded agreement implementation uses listening and sending sets and provides guarantees among core set processes, as specified below.

\begin{definition}[Core-set graded agreement]
Each process $p_i$ in core set $C_{4k+1}$ receives input $v_i$ and outputs some $(v,g)$ where $g \in \{0,1\}$ is a binary grade, such that:
(1)~\textbf{Validity}: if all honest processes in $C_{4k+1}$ start with the same input value $v$, then all honest processes in $C_{4k+1}$ output $(v,1)$;
(2)~\textbf{Graded consistency}: if some honest process in $C_{4k+1}$ outputs $(v,1)$, then every honest process in $C_{4k+1}$ outputs $(v,g)$ for some $g\in\{0,1\}$; and
(3)~\textbf{Termination}: every honest process in $C_{4k+1}$ eventually outputs a pair $(v,g)$ and terminates.
\end{definition}

The graded agreement protocol is executed after $\pi_i$ and $Send_i$ are computed, following the classical two-round structure of graded consensus~\cite{feldman1988optimal}.

The protocol uses two synchronous communication rounds, the \emph{exchange} and \emph{decide} rounds. In the exchange round, each core process sends its value to all processes in its sending set. It then collects exchange messages from all processes in its listening set. If it observes at least $3k+1$ exchange messages from its listening set with the same value $\hat v$, it updates its own value to $\hat v$ and propagates that value again to its sending set in the decide round. Whether or not it sent a decide message itself, it then collects decide messages from its listening set, and uses these to arrive at a final decision.%, depending on its confidence in the decision value. 
In particular, it makes one of three decisions: (1) if it sees at least $3k+1$ copies of the same decision value from its listening set, it decides that value with grade 1, (2) otherwise, if it sees the same decision value from at least $k+1$ listening set processes, then it decides that value but with grade 0, and finally, (3) if it does not see any decision value at least $k+1$ times, it decides on its own value, again with grade 0.
In both rounds, process $p_i$ counts only messages received from processes in its local listening set $\Pi_i^{4k+1}$; all other messages are ignored.

\begin{figure}
\begin{lstlisting}
@$p_i$@'s local variables:
    @$a_i$@       // prediction string
    @$Send_i$@    // sending set

gradedAgreement(v):
    myval = v
    if @$p_i\in C_{4k+1}$@: 
        send <EXCHANGE, myval> to all processes in @$Send_i$@
    if the same EXCHANGE value @$\hat v$@ received at least @$3k+1$@ times from processes in @$\Pi_i^{4k+1}$@:
        myval = @$\hat v$@
        send <DECIDE, myval> to all processes in @$Send_i$@
    R = set of received DECIDE messages from processes in @$\Pi_i^{4k+1}$@
    if the same value @$\hat v$@ received appears in R at least @$3k+1$@ times:
        return @$\hat v, 1$@
    else if the same value @$\hat v$@ appears in R at least @$k+1$@ times:
        return @$\hat v, 0$@
    else:
        return myval, 0
\end{lstlisting}
\caption{Graded Agreement with Prediction}
\label{alg:graded}
\end{figure}

The correctness proof is the standard two-round graded-consensus argument,
adapted to the local sending and listening sets. The replacement for
all-to-all communication is Lemma~\ref{lem:overlap}: whenever an honest
true-prefix process sends to its sending set, every honest core process that
listens to it receives the message. Thus every honest core process receives
messages from at least \(3k+1\) processes in
\(\Pi_i^{4k+1}\cap\hat\Pi^{4k+1}\). The proof appears in
Appendix~\ref{app:graded}.

\begin{theorem}[Core-set graded agreement correctness]
\label{prop:graded-correctness}
Algorithm~\ref{alg:graded} satisfies validity, graded consistency, and
termination for honest processes in \(C_{4k+1}\).
\end{theorem}

\subsection{Maintained Sets and Counters}\label{sec:PK-sets}
Each core process $p_i\in C_{4k+1}$ maintains two local sets and two types of per-process counters, updated in each round $r$:
\begin{itemize}[leftmargin=*, nosep]
    \item \textbf{Candidate Set ($Cand_i^{(r)}$)}: Valid candidate leaders for round $r$, initialized to $Cand_i^{(1)}=\{p_j : cc_i[j]=1\}$.
    \item \textbf{Ignore Set ($Ignore_i^{(r)}$)}: Monotonically non-decreasing set of identified Byzantine processes, initialized to $Ignore_i^{(0)} = \emptyset$.
    \item \textbf{Support Counters}: $support_i^{(r)}(j) = |\{x\in (\Pi_i^{4k+1}\setminus Ignore_i^{(r)}) : j\in Cand_x^{(r)}\}|$, tracking non-ignored listening processes including $p_j$, initialized with $support_i^{(0)}(j)=4k+1$.
    \item \textbf{Front-Runner Counters}: $front_i^{(r)}(j) = |\{x\in (\Pi_i^{4k+1}\setminus Ignore_i^{(r)}) : j=\min Cand_x^{(r)}\}|$, tracking processes where $p_j$ is the minimum-ID candidate, initialized with $front_i^{(0)}(j) = 0$ (no minimum if $Cand_x^{(r)}=\emptyset$).
\end{itemize}

\subsection{The protocol}\label{sec:PK-protocol}

\begin{figure}[!t]
\begin{lstlisting}
local variables for process @$p_i$@: @$Send_i = \emptyset$@, input @$v_i$@, prediction @$a_i$@

Value PCLA():
    if @$\pi_i(i) > 5k+1$@: //not part of core
        wait until receive at least 3k+1 DECIDE(v) messages from processes in @$\Pi_i^{4k+1}$@
        decide v and terminate
    initialize()
    Cand = generateCandidates()
    stop = 2; Ignore = @$\emptyset$@; support[1..n] = 4k+1; front[1..n] = 0
    while true:
        if stop <= 0: return @$v_i$@ 
        if stop == 1: stop -= 1
        (newval, g) = gradedAgreement(@$v_i$@)
        (newg, gg) = gradedAgreement(g)
        if (newg, gg) == (1,1):
            decide newval and broadcast DECIDE(newval) to all processes
            stop -= 1
        maxg = max(g, newg)
        send Cand to all processes in @$Send_i$@
        RCands = set of tuples (x, @$Cand_x$@) where @$p_x \in \Pi_i^{4k+1}$@ and @$Cand_x$@ is received candidate set from @$p_x$@
        for each (x, @$Cand_x$@) in RCands:
            if @$\exists p_j \in Cand_x$@ such that support[j] @$\leq$@ k:
                add @$p_x$@ to Ignore //ignore procs that failed cleanup rule
        old_front[1...n] = front[1...n]; 
        support[1..n] = 0; front[1..n] = 0 //update support and front counters
        for each (x, @$Cand_x$@) in RCands:
            if @$p_x \in Ignore$@: continue // Not counting ignored processes
            for each @$p_j \in Cand_x$@: support[j]++
            if @$Cand_x \neq \emptyset$@: front[@$\min Cand_x$@]++
        if front[i] @$\geq$@ 3k+1:
            send LEADER(newval) to all processes 
        UsedLeaders = @$\emptyset$@
        upon receiving LEADER(v) message from @$p_j$@:
            add @$p_j$@ to UsedLeaders 
            if @$p_j \in$@Cand && support[j] @$\geq$@ 3k+1 && front[j] @$\geq$@ k+1 && maxg == 0:
                newval = v //accept @$p_j$@ as leader
        RemoveSet = {@$p_j$@:support[j]< 3k+1} @$\bigcup$@ UsedLeaders @$\bigcup$@ {@$p_j$@:front[j]@$\geq$@3k+1 and old_front[j]@$\geq$@3k+1}
        Cand = Cand @$\setminus$@ RemoveSet @\label{line:cleanup}@
        @$v_i$@ = newval
\end{lstlisting}
    \caption{The Predictive Core-Leader Agreement (PCLA) Algorithm}
    \label{alg:phaseking}
\end{figure}

The pseudocode for PCLA appears in Algorithm~\ref{alg:phaseking}, with its complete formal proof given in Appendix~\ref{app:phaseking}.
The protocol proceeds as follows. Each core process begins with the two setup phases: it sends pings to build $Send_i$, and exchanges listening sets to construct its initial candidate set $Cand_i^{(1)}$.

Each round of the main loop begins with two consecutive calls to core-set graded agreement; once with $p_i$'s current value, to receive $newval_i, g_i$ as output, and the other with the outputted grade from the previous graded agreement instance (i.e., inputting $g_i$ to the second instance). If the second instance outputs $(1,1)$, $p_i$ then decides the output of the first instance $newval_i$, and broadcasts its decision to all processes. 
This double graded agreement scheme allows safe early stopping -- if a process receives a double grade-1, it can decide immediately and terminate after one more round, with the knowledge that all other processes have necessarily adopted its decision value and will decide in the next round as well. Importantly, a leader's value is only adopted in a given round if the maximum grade of the two graded agreement instances in that round is $0$. Graded agreement therefore also serves as a validity check, preventing a Byzantine leader from injecting an invalid value.

We formalize this intuition in the following lemma that establishes the safety of the double grade-1 termination policy described above.

\begin{lemma}
\label{lem:DecideInOneMoreRound}
If some honest core process first decides value \(v\) in round \(r\), then all honest core processes decide by the end of round \(r+1\).
\end{lemma}

The remainder of the protocol is to ensure that this eventually occurs; in particular, that we eventually have a good round in which some honest process is guaranteed to decide. For this purpose, processes trade candidates and use this information to attempt to converge on a value through the selection of a leader.

In more detail, after the graded agreement instances have been run, process $p_i$ now sends its candidate set $Cand_i^{(r)}$ to all processes in its sending set. Once it receives $Cand_j^{(r)}$ from all processes $p_j$ in its listening set who sent their candidate sets in this round, $p_i$ uses these candidate sets to (a) determine who should be leader in this round, and (b) update its own candidate set for the next round. It also uses this information to detect Byzantine reports and eliminate senders of such reports from its consideration going forward. After receiving the candidate sets, $p_i$ checks whether it should act as leader in this round, by checking whether it is the front runner in at least $3k+1$ received candidate sets. If so, it sends a $\langle LEADER, newval_i\rangle$ to all processes.
Importantly, if $p_i$ does not believe itself to be the leader, it uses different conditions to determine whether to accept a leader message if it receives one. In particular, $p_i$ accepts a leader message from $p_j$ if $p_j$ is in its candidate set $Cand_i^{(r)}$ and (a) appears in the candidate set of at least $3k+1$ processes (i.e., $support_i^{(r)}(j) \geq 3k+1$, and (b) is the front runner of at least $k+1$ processes (i.e., $front_i^{(r)}(j) \geq k+1$). If a leader is accepted by $p_i$, then $p_i$ adopts its value if the maximum grade it got from its two graded agreement instances in this round was $0$.
We show that this acceptance of a leader ensures uniqueness per process per round. More formally,

\begin{lemma}[Local uniqueness of accepted leader]
\label{lem:single king promised}
In every round, each honest process accepts at most one leader proposal.
\end{lemma}

\begin{proof}
Suppose $p_i$ accepts $j < \ell$, requiring $support_i^{(r)}(j)\ge 3k+1$ and $front_i^{(r)}(\ell)\ge k+1$. Since any candidate set with minimum $\ell$ cannot contain $j$, the two counts are disjoint. This requires at least $(3k+1)+(k+1) > 4k+1$ distinct senders in $\Pi_i^{4k+1}$, contradicting $|\Pi_i^{4k+1}|=4k+1$.
\end{proof}

\begin{lemma}\label{lem:king value consistency}
    If some honest process $p_k$ acts as leader in round $r$, and some honest process $p_i$ has $maxg=1$ in round $r$, then $p_k$ sends a $LEADER(v)$ message with the same value $v$ that $p_i$ has as its $newval$ in round $r$.
\end{lemma}

\begin{proof}
    This is a simple consequence of the correctness of the graded agreement protocol. 
    For $p_i$ to have $maxg = 1$ in round $r$, it must have gotten grade 1 in at least one of the graded agreement instances in round $r$. By validity, if the second invocation
    returns value \(1\), then some honest core process input \(1\) to it, and hence some honest core process received grade \(1\) in the first invocation. By the graded consistency of the first instance, all honest processes received the same value \texttt{newval} in the first instance. Thus, since $p_i$ and $p_k$ are both honest, and $p_k$ sends its \texttt{newval} value in its leader message, $p_i$ and $p_k$ have the same value.
\end{proof}

Note that there is an asymmetry in leader selection; $p_i$ needs more certainty to act as leader than to accept another process that claims to be leader. This asymmetry turns out to be the balance that is needed to ensure that a good leader is eventually chosen and listened to by all honest core processes despite their differences in listening and candidate sets. In particular, we show that eventually, there is a round in which some honest process is the accepted leader for all honest core processes.

However, for the latter condition to hold, we must carefully prune the candidate sets in each round to ensure progress towards this goal; in particular, any processes that turn out \emph{not} to be $p^\star$ must be removed from the candidate sets to ensure that $p^\star$ is eventually the front runner in sufficiently many candidate sets.
Thus, the candidate set is pruned in a clean-up phase that removes: (1)~any process $p_j$ with $support_i^{(r)}(j) \le 3k$, as it can never be accepted by $p_i$ in future rounds; (2)~any process $p_j$ that sent a leader message in round $r$; and (3)~any process $p_j$ with $front_i^{(r)}(j) \ge 3k+1$ across two consecutive rounds, since an honest leader would have already acted.

Given these clean-up rules, $p_i$ can also detect Byzantine behavior of processes that did not properly execute their clean up from the previous round. Thus, if any process $p_\ell$ lists in its candidate set a process $p_j$ that was supported by $k$ or fewer processes in the previous round, then $p_\ell$ is added to $p_i$’s ignore set. Intuitively, $p_i$ knows that any honest process $p_\ell$ would have removed such a $p_j$ from its candidate set before reporting $Cand_\ell^{(r)}$, and therefore, if $p_\ell$ did not do so, it must be Byzantine. This ignore-set update can be executed immediately upon receiving the candidate sets in round $r$, and does not need to wait until the end of the round.
We show that an honest process never places an honest process in its ignore set.

\begin{lemma}[Ignore set safety]
\label{lem:ignore_set_safety}
For every honest process \(p_i\in C_{4k+1}\) and every round \(r\ge 0\), $Ignore_i^{(r)}\cap \mathcal H=\emptyset$.
\end{lemma}

\begin{proof}[Proof sketch]
The ignore rule adds a process only if it reports a candidate whose previous
support was at most \(k\). We show that an honest process can never do this; if an honest process \(p_h\) keeps a candidate \(p_j\) into the next
round, then \(p_h\) must have seen support at least \(3k+1\) for \(p_j\) in the
previous round. At least \(2k+1\) of these supporters lie in
\(\hat\Pi^{4k+1}\), and for any other honest core process $p_i$, its listening set's overlap with these is at least $k+1$.
\end{proof}

 To prove that the algorithm terminates, we rely on the fact that honest core processes have a common honest candidate (as established by Lemma~\ref{lem:common honest}), and show that progress is made toward such a candidate being the front runner for enough processes. For that purpose, denote the lowest-ID candidate that all honest core processes have in common by $p^\star = \min\left(\hat\Pi^{4k+1} \cap \bigcap_{p_i\in C_{4k+1}\cap \mathcal{H}} Cand_i^{(1)}\right)$.
 We show that bad candidates with lower IDs than $p^\star$ are regularly eliminated from honest processes' candidate sets.

 \begin{lemma}[Three-round elimination of the global honest minimum]
\label{lem:Cand_gettingSmaller_for_all_honest}
     Let $m^{(r)}$ be the minimum candidate across all honest core processes' candidate sets in round $r$. Then $m^{(r)} \not\in Cand_i^{(r+3)}$ for every honest core process $p_i$.
 \end{lemma}

 \begin{proof}[Proof sketch]
Let \(m=m^{(r)}\). By definition of \(m\), no honest
core process has any candidate \(q<m\) at the beginning of round \(r\). Thus,
from the point of view of any honest core process \(p_i\), any report containing
such a \(q\) can only come from the at most \(k\) wrongly-predicted-correct processes in
\(\Pi_i^{4k+1}\setminus\hat\Pi^{4k+1}\). Therefore \(q\) has support at most
\(k\) at \(p_i\).

Now consider round $r+1$. If some process $p_j$ reports a candidate
\(q<m\), then \(p_i\) detects that $p_j$ failed to remove a candidate
whose previous support was \(\leq k\), and adds $p_j$ to the ignore set before
recomputing support and front. Consequently, after the ignore-set
update in round \(r+1\), no active candidate set seen by \(p_i\) contains a
candidate smaller than \(m\).

At this point, $m$ will be removed from $p_i$'s candidate set within two rounds due to the cleanup rules -- either its support diminishes to below $3k+1$ processes, or it acts as LEADER, or it remains the front runner of at least $3k+1$ candidate sets for two rounds in a row. 
\end{proof}

To bound the amount of time it takes for a good candidate to emerge, we define the set of \emph{blocking candidates}, or processes with a lower ID than $p^\star$ that appear in the candidate set of some honest core process. Intuitively, these are the $m$s that we must spend rounds to remove from the candidate sets.
\[
BlockingCands
=
\{\,q<p^\star : \exists p_j\in C_{4k+1}\cap H \text{ such that } q\in Cand_j^{(1)}\,\}.
\]

\begin{lemma}[Bound on blocking candidates]
\label{lem:pk-blocking-k}
The number of blocking candidates satisfies
\(|BlockingCands|\le 8s\).
\end{lemma}

\begin{proof}
We show that every blocking candidate is either a classified false negative or
a classified false positive. Fix \(q\in BlockingCands\). If
\(q\in\hat\Pi^{4k+1}\), then \(q\) is an honest core process. Since
\(q<p^\star\), and \(p^\star\) is the minimum-ID honest core process classified
positively by every honest core process, \(q\) cannot be classified positively
by all honest core processes. Hence some honest core process rejects \(q\), so
\(q\in\mathrm{FalseNeg}_{4k+1}\).

Otherwise, \(q\notin\hat\Pi^{4k+1}\). Since \(q\in BlockingCands\), it appears
in the initial candidate set of some honest core process, equivalently it is
classified positively by that process. Hence
\(q\in\mathrm{FalsePos}_{4k+1}\).
Thus \(BlockingCands\subseteq
\mathrm{FalseNeg}_{4k+1}\cup\mathrm{FalsePos}_{4k+1}\). The claim follows from
Lemma~\ref{lem:classification-error-budget}.
\end{proof}

\begin{definition}[Good leader round]
A round $r$ is a \emph{good leader round} if, after the two graded-consensus invocations of round $r$, there exists an honest core process $p^\star$ such that:
\begin{enumerate}[nosep]
    \item $p^\star$ sends $\textsf{LEADER}(v)$ in round $r$;
    \item every honest core process with $maxg=0$ accepts the value sent by $p^\star$;
    \item every honest core process with $maxg=1$ has
    $newval=v$.
\end{enumerate}
\end{definition}

\begin{lemma}[Existence of a good leader]
\label{lem:good-king-exists}
Let \(p^\star\) be the minimum-ID honest core process that is classified
positively by every honest core process. Then either some honest core process
decides before \(p^\star\) becomes a good leader, or there exists a round in which
\(p^\star\) is a good leader.
\end{lemma}

\begin{proof}[Proof sketch]
The proof relies on two technical facts, proved formally in
Appendix~\ref{app:phaseking}. First, as long as no honest core process
ever sees \(p^\star\) as a strong front-runner, \(p^\star\) is never removed
from any honest core candidate set. Second, since the number of blocking
candidates below \(p^\star\) is bounded by Lemma~\ref{lem:pk-blocking-k}, and
Lemma~\ref{lem:Cand_gettingSmaller_for_all_honest} eliminates the global honest
minimum within a constant number of rounds, eventually some honest core process
must see \(p^\star\) as a strong front-runner.
We show that $p^\star$ remains in the candidate set of all honest processes until enough of them see it as the front runner. Once that happens, $p^\star$ can act as leader, and by Lemma~\ref{lem:king value consistency}, this becomes a good leader round.
\end{proof}
A good leader round synchronizes the values of all honest core processes: processes
with \(\texttt{maxg}=0\) adopt the leader value, while processes with
\(\texttt{maxg}=1\) already hold that value by definition of a good leader round.
Therefore, by validity of the two graded-agreement invocations, all honest core
processes decide in the next round. 

Until now, we've shown that all decisions are safe, and that eventually, a decision among the core set processes is reached (since, if a decision does not happen earlier, eventually a good leader round occurs). 
We now wrap up the proof with a few final lemmas that show how a decision gets disseminated to the rest of the processes and formalizing the above statements to show the desired agreement, validity, and termination properties.

\begin{lemma}[Dissemination to non-core processes]
\label{lem:pk-dissemination}
Assume that all honest core processes output the same value $v$ and broadcast $\textsf{DECIDE}(v)$ to all processes. Then every honest process $p_i\notin C_{4k+1}$ eventually decides $v$.
\end{lemma}

\begin{proof}
Let $p_i\notin C_{4k+1}$ be honest. By Lemma~\ref{lem:pi hat intersect}, $|\Pi_i^{4k+1}\cap \hat\Pi^{4k+1}|\ge 3k+1$.
By Lemma~\ref{lem:C size}, every process in $\hat\Pi^{4k+1}$ belongs to $C_{4k+1}$. Therefore, among the processes in $\Pi_i^{4k+1}$, at least $3k+1$ are honest core processes. By assumption, all of them broadcast $\textsf{DECIDE}(v)$. Hence $p_i$ eventually receives $\textsf{DECIDE}(v)$ from at least $3k+1$ processes in $\Pi_i^{4k+1}$ and decides $v$.
\end{proof}

\begin{theorem}[Correctness]
\label{thm:pk-correctness}
The Predictive Core-Leader Agreement protocol satisfies validity, agreement, and termination.
\end{theorem}
\section{Complexity}
\label{sec:complexity}

We now bound the round and honest-message complexity. Throughout this section,
\(f\) denotes the actual number of Byzantine processes, \(k\) is the local
prediction-error guarantee, and \(s\le k\) is the actual maximum prediction
error of any honest process in the execution.

\begin{theorem}[Complexity]
\label{thm:pk-complexity}
The Predictive Core-Leader Agreement (PCLA) protocol reaches a decision after \(O(s+1)\)
iterations of the main loop. Consequently, its round complexity is \(O(s+1)\), and
\(O(k)\) in the worst case. Its honest-message complexity is
\(O((f+k)ks+nk)\). In particular, in the worst case this is
\(O((t+k)k^2+nk)\).
\end{theorem}

\begin{proof}
By Lemma~\ref{lem:C size}, there are at most \(6k+1\) honest core processes.
Moreover, any honest process sends only to its sending set, whose size is at
most \(f+6k+1\): the sending set contains at most all honest core processes and
the \(f\) Byzantine processes. Thus, whenever all honest core processes send one
message to their sending sets, the cost is \(O((f+k)k)\).

The setup phase consists of sending-set construction and candidate generation.
Sending-set construction costs \(O(k^2)\) messages, since each honest core
process sends to \(4k+1\) processes. Candidate generation costs
\(O((f+k)k)\) messages by the sending-set bound above. Thus the total setup cost
is \(O((f+k)k)\) messages and \(O(1)\) rounds.

Each main-loop iteration contains two graded-agreement invocations and one
candidate-set transmission step. Each of these costs \(O((f+k)k)\) honest
messages and \(O(1)\) rounds, so one iteration costs \(O((f+k)k)\) messages and
\(O(1)\) rounds. By the progress argument, after \(O(s+1)\) iterations either a
decision has already occurred or a good synchronizing round occurs, which leads to
decision within a constant number of additional iterations.

Finally, each honest core process broadcasts \(\textsf{DECIDE}\) at most once
and sends \(\textsf{LEADER}\) at most once before being removed as a used leader.
Since there are \(O(k)\) honest core processes, all such broadcasts contribute
\(O(nk)\) messages in total. Therefore the total honest-message complexity is $O((f+k)k) + O((s+1)(f+k)k) + O(nk) = O((f+k)ks+nk)$, where the additive \(O((f+k)k)\) term is absorbed by \(O(nk)\). Since \(s\le k\)
and \(f\le t\), the worst-case message bound is \(O((t+k)k^2+nk)\), and the
worst-case round bound is \(O(k)\).
\end{proof}

\begin{corollary}[Prefix-sensitive complexity]
\label{cor:pk-prefix-sensitive-complexity}
Let \(\sigma\) be the prefix-sensitive prediction error defined in
Remark~\ref{rem:sigma-defintion}. Then the protocol reaches a decision after
\(O(\sigma+1)\) main-loop iterations, has round complexity \(O(\sigma+1)\),
and has honest-message complexity \(O((f+k)k\sigma+nk)\).
\end{corollary}

\section{Lower Bounds}\label{sec:lower-bounds}

In this section, we prove lower bounds for the local $k$-error prediction model. 
As in the classical Dolev--Reischuk lower bound~\cite{dolev1985bounds}, and as in prior work on Byzantine Agreement with predictions, we state the main message lower bound for Byzantine broadcast. 
Byzantine broadcast assumes a designated sender and requires that all honest processes deliver the same value, and that if the sender is honest then all honest processes deliver the sender's input value. 
The standard reduction from Byzantine broadcast to Byzantine agreement adds only an initial sender-to-all communication step, and therefore an $\Omega(nk)$ lower bound for broadcast also gives the same asymptotic lower bound for Byzantine Agreement when $k \ge 1$.

We first show that deterministic synchronous Byzantine broadcast in the local $k$-error prediction model requires $\Omega(nk)$ honest messages even in an all-honest execution in which all processes receive the same perfectly accurate prediction vector. 
The key to the argument is that, although the hard execution itself has no prediction errors, the protocol must remain correct on executions in which up to $k$ processes are falsely predicted to be honest. 
We then show a black-box reduction from standard Byzantine broadcast, showing that any deterministic lower bound for the classical problem carries over whenever the number of faults does not exceed the local prediction budget.
  
\subsection{An $\Omega(nk)$ Message Lower Bound}

We begin with the direct lower bound. 
Even if the actual execution is all-honest and all processes receive the same perfectly accurate prediction vector, any deterministic synchronous Byzantine broadcast protocol in the local $k$-error prediction model must send $\Omega(nk)$ honest messages in some such execution.

\begin{theorem}
\label{thm:msg-lb-local-k}
Assume $t \ge k \ge 1$. 
For every deterministic synchronous algorithm that solves Byzantine broadcast with up to $t$ Byzantine faults in the local $k$-error prediction model, there exists an all-honest execution in which all processes receive the same perfectly accurate prediction vector, and in which honest processes send $\Omega(nk)$ messages.
\end{theorem}

\begin{proof}[Proof Sketch.]
We adapt the classical indistinguishability technique of Dolev and Reischuk~\cite{dolev1985bounds} for the prediction-free $\Omega(nt)$ lower bound. Consider two all-honest executions $E_0, E_1$ with sender inputs $0, 1$ and identical accurate predictions $(1,\dots,1)$, and sender $p_1$.

We define the \emph{touch count} of a process $p_j$, $\mathrm{touch}(p_j)$, to be the total messages incident to $p_j$ (sent to or from $p_j$) across both executions. If in both executions $E_0, E_1$, there are $< nk/8$ honest messages, then by a loose counting argument, we can show that there is at least one process $p_i$ other than the sender such that $touch(p_i) < k$.
Let $\Lambda$ be the set of processes that exchange at least one message with $p_i$ in $E_0$ or $E_1$, so $|\Lambda| \le \mathrm{touch}(p_i) < k \le t$. We construct a third execution $E_{\mathrm{bad}}$ in which all processes in $\Lambda$ are Byzantine, all other processes (including $p_i$) are honest, the sender $p_1$ has input $0$ if $p_1 \notin \Lambda$, and every honest process keeps the all-ones prediction vector $(1,\dots,1)$. The Byzantine processes in $\Lambda$ behave as in $E_0$ toward all honest processes other than $p_i$, and as in $E_1$ toward $p_i$. Consequently, honest process $p_i$ cannot distinguish $E_{\mathrm{bad}}$ from $E_1$ and decides $1$, whereas all other honest processes cannot distinguish $E_{\mathrm{bad}}$ from $E_0$ and decide $0$, violating agreement. The full proof is in Appendix~\ref{app:lowerbound}.
\end{proof}

Furthermore, any classical deterministic lower bound for synchronous Byzantine broadcast carries over whenever the number of faults does not exceed the local uncertainty bound ($\ell \le \min\{t,k\}$): an adversary can simply induce an execution where all honest nodes receive the all-ones vector $(1,\dots,1)$, incurring at most $\ell \le k$ false positives and zero false negatives. Consequently, any protocol in our model inherits the standard message lower bounds on this execution family, and the worst-case round complexity is at least $\min\{t,k\}+1$ (formal statements and proofs in Appendix~\ref{app:lowerbound}).

\subsection{Common Prediction Vectors and the Price of Disagreement}
\label{sec:common-vector}

The \(s\)-dependent part of our protocol is the cost of reconciling different
honest local prediction views. This cost disappears when all honest processes
receive the same prediction vector.

Suppose all honest processes receive the same vector \(a\) with
\(d_H(a,\hat a)\le k\), and assume \(n-t\ge 3k+1\). Let \(C\) be the first
\(3k+1\) processes in the ordering induced by \(a\). Then \(C\) contains at most
\(k\) Byzantine processes. Hence the processes in \(C\) can simply run any
Byzantine Agreement protocol \(\mathcal A\) for \(3k+1\) processes tolerating
\(k\) Byzantine faults, and then disseminate the decision to all processes:
each process outside \(C\) decides after receiving \(k+1\) matching
\(\textsf{DECIDE}\) messages from members of \(C\).

This gives Byzantine Agreement with message complexity $M_{\mathcal A}(3k+1,k)+O(nk)$
and, in the synchronous model, round complexity 
$R_{\mathcal A}(3k+1,k)+1$.\\
Thus, using a standard \(O(k^2)\)-message committee protocol, the total message
complexity is \(O(nk)\), matching the \(\Omega(nk)\) lower bound up to constants.
This illustrates that the extra $O((f+k)ks)$ term in PCLA is
the price of resolving disagreement between different honest prediction views. We therefore conjecture that the lower bound can be tightened, but this would require a different technique that explicitly relies on an execution in which processes receive different prediction vectors.

\section*{Acknowledgements}
This work is partially supported by ISF grant number 3673/25. 

\subsection*{AI Disclosure} We used ChatGPT to assist with positioning the paper, as well as cleaning up the presentation of the technical sections. The tool materially affected Sections~\ref{sec:intro} and~\ref{sec:related}. Additionally, ChatGPT helped tighten the analysis of the algorithm's resilience, from $n\geq t+ 5k +1$ to $n\geq t+ 4k +1$. The authors verified the correctness and originality of all content including references.

\bibliographystyle{plainurl}
\bibliography{references}

\newpage
\appendix

\section{Listening and Core Set Honest Overlap Proofs}\label{app:overlap}

\begin{lemma}
\label{lem:pi hat intersect-app}
For any honest process $p_i$ and any $x\le n-t$,
\[
|\Pi_i^x \cap \hat\Pi^x| \ge x-k.
\]

\end{lemma}
\begin{proof}
Let $a_i^{FN}$ be an intermediate prediction vector derived from $a_i$ by setting all False positive bits to $0$. Thus, $a_i^{FN}[j]=1$ implies that $p_j$ is honest $(\hat a[j]=1)$. Let $k_{FN}\le k$ be the number of False Negative errors in $a_i^{FN}$ (and $a_i$). Let $\pi_{i-FN}$ be the implied permutation and define $\Pi_{i-FN}^x$ similarly.\\ Since $a_i^{FN}$ is ``sparser" than the ground truth $\hat a$, for any honest process $p_j$ that is predicted correctly, its rank in the mediator permutation $\pi_{i-FN}$ is: $ \hat\pi (j)=\sum_{r=1}^{j}\hat{a}[r]\ge \sum_{r=1}^{j}a_i^{FN}[r]=\pi_{i-FN}(j)$. This implies that any process in the first $x$ positions of $\hat\pi$ that is also predicted as honest by $a_i^{FN}$ will have a rank $\le x$ in $\pi_{i-FN}$. Thus,  $|\Pi_{i-FN}^x\cap \hat\Pi^x|\ge x-k_{FN}$. \\
Next, let $k_{FP}=d_H(a_i,a_i^{FN})$ be the number of False Positive errors in $a_i$, where $k_{FN}+k_{FP}= k$. Because $a_i^{FN}[j]=1 \rightarrow a_i[j]=1$, the rank of any honest process $p_j$ can only increase when transitioning from $\pi_{i-FN}$ to $\pi_i$ : $\pi_i(j) =  \sum_{r=1}^ja_i[r]\le \pi_{i-FN}(j) + k_{FP}$. Crucially, the rank increases by at most $k_{FP}$. Since $\Pi_{i-FN}^x\cap \hat\Pi^x\subseteq H$ then we can deduce that $|\Pi_i^x\cap(\Pi_{i-FN}^x\cap \hat\Pi^x)|\ge \underbrace{|\Pi_{i-FN}^x\cap \hat\Pi^x|}_{\ge x - k_{FN}} - k_{FP}\ge x-k$. Because $(\Pi_{i-FN}^x\cap \hat\Pi^x)$ is a subset of $\hat\Pi^x$, we conclude $|\Pi_i^x\cap \hat\Pi^x|\ge x-k $.
\end{proof}

\begin{lemma}[Honest core size]\label{lem:C size-app}
For any $x\le n-t$, the set $C_x$ contains all the first $x$ processes in $\hat\pi$, i.e.\ $\hat\Pi^x\subseteq C_x$, and 
$|C_x\cap \mathcal H|\le x+2k$.
\end{lemma}
\begin{proof}
Every honest process predicts itself as honest, therefore $\pi_i(i)\le\hat\pi(i)+k$. If $\hat\pi(i)\le x$, then $\pi_i(i)\le x+k \implies p_i\in C_x$.

For the size bound on $C_x\cap\mathcal{H}$, let $p_i\in C_x\cap\mathcal{H}$. Because $p_i$ predicts itself as honest ($a_i[i]=1$), its local rank satisfies $\pi_i(i) = \hat\pi(i) + \mathrm{FP}_i^{\le i} - \mathrm{FN}_i^{\le i} \ge \hat\pi(i) - k$, where $\mathrm{FN}_i^{\le i}$ is the number of false negatives with index at most $i$. Since $p_i\in C_x$, we have $\pi_i(i)\le x+k$, so $\hat\pi(i) - k \le x+k$, which implies $\hat\pi(i)\le x+2k$. Therefore, every honest process in $C_x$ belongs to $\hat\Pi^{x+2k}$, proving $|C_x \cap \mathcal{H}| \le \min(x+2k, n) \le x+2k$.
\end{proof}

\begin{lemma}[Listening set overlap]
\label {lem:coreset-coherence-appendix}
Let $p_i$, $p_j$ be honest processes, and let $x\le n-t$. Then $|\Pi^x_i \cap\Pi_j^x \cap \mathcal{H}| \ge x-2k$.
\end{lemma}
\begin{proof}
Since $(\Pi_i^x \cap \hat\Pi^x) \cup (\Pi_j^x \cap \hat\Pi^x) \subseteq \hat\Pi^x$ therefore from the inclusion-exclusion principle: 
$${|(\Pi_i^x \cap \hat\Pi^x)| + |(\Pi_j^x \cap \hat\Pi^x)| - |(\Pi_i^x \cap \hat\Pi^x) \cap (\Pi_j^x \cap \hat\Pi_x)|\leq |\hat\Pi^x|}$$
By \autoref{lem:pi hat intersect-app}:
$${\underbrace{|(\Pi_i^x \cap \hat\Pi^x)|}_{\text{$\ge x -k$}} + \underbrace{|(\Pi_j^x \cap \hat\Pi^x)|}_{\text{$\ge x -k$}} - \underbrace{|\hat\Pi^x|}_{\text{$x$}} \leq |(\Pi_i^x \cap \hat\Pi^x) \cap (\Pi_j^x \cap \hat\Pi^x)|}$$
Therefore $ x-2k \leq |(\Pi_i^x \cap \hat\Pi^x) \cap (\Pi_j^x \cap \hat\Pi^x)|$ and since $$(\Pi_i^x \cap \hat\Pi^x) \cap (\Pi_j^x \cap \hat\Pi^x) \subseteq \Pi_i^x \cap \Pi_j^x$$ it implies $|\Pi_i^x \cap \Pi_j^x| \ge x-2k$. Since $x\le n-t$ then $\hat\Pi^x \subseteq \mathcal{H}$ and therefore all of these processes are honest.
\end{proof}

The following corollary shows that there is significant overlap in the listening sets of honest processes, and in particular, that there is a quorum of size at least $2k+1$ of honest processes of overlap between any two honest processes' listening sets.

\begin{corollary}[Listening-set overlap]
\label{cor:listening-overlap-app}
For any two honest processes \(p_i,p_j\),
\(|\Pi_i^{4k+1}\cap \Pi_j^{4k+1}\cap\mathcal H|\ge 2k+1\).
\end{corollary}

\begin{proof}
By Lemma~\ref{lem:pi hat intersect-app}, each of \(\Pi_i^{4k+1}\) and
\(\Pi_j^{4k+1}\) contains at least \(3k+1\) processes from
\(\hat\Pi^{4k+1}\). Since \(|\hat\Pi^{4k+1}|=4k+1\), their overlap inside
\(\hat\Pi^{4k+1}\) has size at least \(2k+1\). All these processes are honest.
\end{proof}

\section{Proofs of Candidate Leader Properties}\label{app:candidates}

\begin{lemma}[Common honest candidate.]
\label{lem:common-honest-appendix}
There exists an honest process $p^\star\in \hat\Pi^{4k+1}$ such that every honest core process classifies it positively:
\[
\forall p_i\in\mathcal H \cap C_{4k+1}:\quad cc_i[p^\star]=1.
\]
\end{lemma}
\begin{proof}
For each honest process $p_r\in \hat\Pi^{4k+1}$, Lemma~\ref{lem:C size} implies that $p_r\in C_{4k+1}$, and Lemma~\ref{lem:pi hat intersect} implies that $|\Pi_r^{4k+1}\cap \hat\Pi^{4k+1}|\ge 3k+1$.

Therefore each honest process in $\hat\Pi^{4k+1}$ votes for at least $3k+1$ processes from $\hat\Pi^{4k+1}$. 
Since there are $4k+1$ honest processes in $\hat\Pi^{4k+1}$, the total number of honest votes cast by processes in $\hat\Pi^{4k+1}$ toward processes in $\hat\Pi^{4k+1}$ is at least $(4k+1)(3k+1)$.
These votes are distributed among the $4k+1$ processes of $\hat\Pi^{4k+1}$. By the pigeonhole principle, there exists a process $p^\star\in \hat\Pi^{4k+1}$ that receives at least $3k+1$ honest votes from processes in $\hat\Pi^{4k+1}$.

Now fix any honest process $p_i$, and let
\[
V=\{p_r\in \hat\Pi^{4k+1} : p_r \text{ votes for } p^\star\}.
\]
Then $|V|\ge 3k+1$. Also, by Lemma~\ref{lem:pi hat intersect}, 
$|\Pi_i^{4k+1}\cap \hat\Pi^{4k+1}|\ge 3k+1$.
Both $V$ and $\Pi_i^{4k+1}\cap \hat\Pi^{4k+1}$ are subsets of $\hat\Pi^{4k+1}$, whose size is $4k+1$. Hence, by inclusion-exclusion,
\[
|V\cap \Pi_i^{4k+1}|
\ge |V| + |\Pi_i^{4k+1}\cap \hat\Pi^{4k+1}| - |\hat\Pi^{4k+1}|
\ge (3k+1)+(3k+1)-(4k+1)=2k+1.
\]
Thus process $p_i$ sees at least $2k+1$ votes for $p^\star$, and therefore sets $cc_i[p^\star]=1$.

Since $p_i$ was arbitrary, every honest process includes $p^\star$ in its classification vector.
\end{proof}

\begin{lemma}[Global classification error bound]
\label{lem:classification-error-budget-appendix}
\[
|\mathrm{FalseNeg}_{4k+1}|+|\mathrm{FalsePos}_{4k+1}|\le 8s.
\]
In particular, this quantity is \(O(s)\); since \(s\le k\), it is at most
\(8k\), and hence \(O(k)\), in the worst case.
\end{lemma}

\begin{proof}
We charge classification errors to bad prefix incidences among the processes in
$\hat\Pi^{4k+1}$.

First fix $p_j\in \mathrm{FalseNeg}_{4k+1}$. Then
$p_j\in\hat\Pi^{4k+1}$, and there exists an honest core process
$p_i\in C_{4k+1}\cap\mathcal H$ such that $cc_i[j]=0$. Hence $p_i$ counted
fewer than $2k+1$ votes for $p_j$. By Lemma~\ref{lem:pi hat intersect}, $|\Pi_i^{4k+1}\cap\hat\Pi^{4k+1}|\ge 3k+1$.
Therefore at least $(3k+1)-2k=k+1$
processes in $\Pi_i^{4k+1}\cap\hat\Pi^{4k+1}$ omitted $p_j$ from their
transmitted prefixes.

Now fix $p_j\in \mathrm{FalsePos}_{4k+1}$. Then
$p_j\notin\hat\Pi^{4k+1}$, and there exists an honest core process
$p_i\in C_{4k+1}\cap\mathcal H$ such that $cc_i[j]=1$. Hence $p_i$ counted at
least $2k+1$ votes for $p_j$. Since $|\Pi_i^{4k+1}\setminus\hat\Pi^{4k+1}|\le k$,
at least $k+1$ of these votes come from processes in $\hat\Pi^{4k+1}$. Each
such vote is an inclusion of $p_j$ in a transmitted prefix although
$p_j\notin\hat\Pi^{4k+1}$.

Thus every classified false negative or false positive can be charged to at
least $k+1$ bad prefix incidences contributed by processes in
$\hat\Pi^{4k+1}$. The charged incidences for different classified errors are
distinct, since each incidence is identified by the sender and the classified
process. Therefore, if
\[
\hat B_{4k+1}
=
\sum_{p_r\in \hat\Pi^{4k+1}}
\left(
|\Pi_r^{4k+1}\setminus \hat\Pi^{4k+1}|
+
|\hat\Pi^{4k+1}\setminus \Pi_r^{4k+1}|
\right),
\]
then 
\[
\left(
|\mathrm{FalseNeg}_{4k+1}|+|\mathrm{FalsePos}_{4k+1}|
\right)(k+1)
\le
\hat B_{4k+1}.
\]

Finally, each process $p_r\in\hat\Pi^{4k+1}$ has at most $s$ prediction errors.
Therefore its local prefix $\Pi_r^{4k+1}$ can differ from
$\hat\Pi^{4k+1}$ in at most $s$ extra processes and at most $s$ missing
processes. Hence each such $p_r$ contributes at most $2s$ to
$\hat B_{4k+1}$, so $\hat B_{4k+1}\le (4k+1)\cdot 2s$.
Combining the two inequalities gives
\[
|\mathrm{FalseNeg}_{4k+1}|+|\mathrm{FalsePos}_{4k+1}|
\le
\frac{(4k+1)2s}{k+1}
\le8s=O(s).
\]
Since \(s\le k\), the worst-case bound is \(O(k)\).
\end{proof}

\section{Correctness of Core-Set Graded Agreement}
\label{app:graded}
\begin{lemma}
    The core-set graded agreement algorithm satisfies validity.
\end{lemma}

\begin{proof}
Suppose all honest processes in $C_{4k+1}$ start with input value $v$. Fix an
honest process $p_i\in C_{4k+1}$.

By Lemma~\ref{lem:overlap}, in the exchange phase $p_i$ receives messages from at
least $3k+1$ honest processes in its listening set. Since all honest core processes
start with value $v$, all these messages contain $v$. Therefore $p_i$ receives at
least $3k+1$ copies of $v$, sets $myval\leftarrow v$, and sends a \textsf{DECIDE} message.

Since $p_i$ was arbitrary, every honest process in $C_{4k+1}$ sends $v$ in the
decide phase. Applying Lemma~\ref{lem:overlap} again, every honest process
$p_i\in C_{4k+1}$ receives at least $3k+1$ copies of $v$ in the decide phase.
Thus every honest process in $C_{4k+1}$ outputs $(v,1)$.
\end{proof}

\begin{lemma}[Decide-phase integrity]
\label{lem:decide-integrity}
If honest processes $p_i,p_j\in C_{4k+1}$ send values $v_i$ and $v_j$,
respectively, in the decide phase, then $v_i=v_j$.
\end{lemma}

\begin{proof}
Suppose honest processes $p_i,p_j\in C_{4k+1}$ send values $v_i$ and $v_j$,
respectively, in the decide phase.

For $p_i$ to send $v_i$ in the decide phase, it must have received at least
$3k+1$ copies of $v_i$ in the exchange phase, all counted from senders in
$\Pi_i^{4k+1}$. By Lemma~\ref{lem:pi hat intersect},
\[
|\Pi_i^{4k+1}\cap\hat\Pi^{4k+1}|\ge 3k+1,
\]
and hence at most $k$ processes in $\Pi_i^{4k+1}$ lie outside
$\hat\Pi^{4k+1}$. Therefore at least $2k+1$ of the senders from which $p_i$
received $v_i$ belong to $\hat\Pi^{4k+1}$.

Similarly, at least $2k+1$ of the senders from which $p_j$ received $v_j$
belong to $\hat\Pi^{4k+1}$. Thus, both $p_i$ and $p_j$ have sets of at least
$2k+1$ honest-prefix senders supporting their respective decide-phase values.
Since both sets are contained in $\hat\Pi^{4k+1}$, which has size $4k+1$, they
must intersect:
\[
(2k+1)+(2k+1) > 4k+1.
\]
Therefore, there is an honest process in $\hat\Pi^{4k+1}$ that sent the value
supporting $p_i$'s decide-phase message and also the value supporting $p_j$'s
decide-phase message during the exchange phase. Honest processes send the same
exchange-phase value to all recipients, so the two supported values must be the
same. Hence $v_i=v_j$.
\end{proof}

\begin{lemma}
    The core-set graded agreement protocol satisfies graded consistency.
\end{lemma}

\begin{proof}
Suppose an honest process $p_i\in C_{4k+1}$ outputs $(v,1)$. Then, in the decide
phase, $p_i$ received at least $3k+1$ copies of $v$ from processes in
$\Pi_i^{4k+1}$.

By Lemma~\ref{lem:pi hat intersect},
\[
|\Pi_i^{4k+1}\setminus \hat\Pi^{4k+1}|\le k.
\]
Therefore at least $2k+1$ of the decide-phase senders from which $p_i$ received
$v$ belong to $\hat\Pi^{4k+1}$.

Now fix any honest process $p_j\in C_{4k+1}$. Again by
Lemma~\ref{lem:pi hat intersect},
\[
|\Pi_j^{4k+1}\cap \hat\Pi^{4k+1}|\ge 3k+1.
\]
Both the $2k+1$ honest-prefix senders of $v$ seen by $p_i$ and the set
$\Pi_j^{4k+1}\cap \hat\Pi^{4k+1}$ are subsets of $\hat\Pi^{4k+1}$, whose size is
$4k+1$. Hence, by inclusion-exclusion, $p_j$ has at least
\[
(2k+1)+(3k+1)-(4k+1)=k+1
\]
of those senders in its listening set $\Pi_j^{4k+1}$.

By the initialization of the sending sets, every such sender sends its decide-phase
message to $p_j$. Thus $p_j$ receives at least $k+1$ copies of $v$ in the decide
phase.

It remains to show that $p_j$ cannot output a different value. By
Lemma~\ref{lem:decide-integrity}, all honest processes that send in the decide
phase send the same value. Since at least one honest process sends $v$ in the
decide phase, no honest decide-phase sender sends a value different from $v$.
Therefore any value different from $v$ can be received by $p_j$ only from
processes in $\Pi_j^{4k+1}\setminus \hat\Pi^{4k+1}$, of which there are at most
$k$ by Lemma~\ref{lem:pi hat intersect}. Hence no value different from $v$ can
appear at least $k+1$ times.

Consequently, $p_j$ outputs $(v,g)$ for some $g\in\{0,1\}$.
\end{proof}

\begin{lemma}
    The core-set graded agreement algorithm satisfies termination.
\end{lemma}
\begin{proof}
The protocol is defined over a fixed sequence of synchronous rounds.
\end{proof}

\section{PCLA Protocol Correctness Proofs}\label{app:phaseking}

\begin{lemma}
\label{lem:DecideInOneMoreRound-app}
If some honest core process first decides value \(v\) in round \(r\), then every
honest core process that decides in round \(r\) decides \(v\), and every honest
core process that does not decide in round \(r\) enters round \(r+1\) with value
\(v\). Consequently, all honest core processes decide by the end of round \(r+1\).
\end{lemma}

\begin{proof}
Let \(p_i\) be an honest core process that decides \(v\) in round \(r\). Then
\(p_i\) outputs \((1,1)\) in the second graded-agreement invocation. Since the
second invocation has binary inputs, not all honest core processes could have
input \(0\); otherwise validity would force output \((0,1)\). Hence some honest
core process input \(1\) to the second invocation, meaning that some honest core
process obtained grade \(1\) in the first invocation.

By graded consistency of the first invocation, all honest core processes obtain
the same value \(v\). By graded consistency of the second invocation, every
honest core process has \(\texttt{maxg}=1\) in round \(r\), so no honest core
process adopts a leader value in that round. Therefore all honest core processes
either decide \(v\) in round \(r\), or enter round \(r+1\) with value \(v\). In
round \(r+1\), validity of the two graded-agreement invocations makes all
remaining honest core processes decide \(v\).
\end{proof}

\begin{lemma}[Ignore set safety]
\label{lem:ignore_set_safety-app}
For every honest process $p_i\in C_{4k+1}$ and every round $r\ge 0$,
\[
Ignore_i^{(r)}\cap \mathcal H=\emptyset.
\]
\end{lemma}
\begin{proof}
We prove the claim by induction on $r$. For $r=0$, it follows from
$Ignore_i^{(0)}=\emptyset$.

First consider the transition from round $0$ to round $1$. Since
$support_i^{(0)}(j)=4k+1>k$
for every candidate $p_j$, the ignore rule cannot add any process to $Ignore_i^{(1)}$.

Thus, $Ignore_i^{(1)}\cap\mathcal H=\emptyset$.

Now assume the claim holds for round $r\ge 1$. Suppose, towards contradiction, that in round
$r+1$ an honest process $p_h$ is added to $Ignore_i$, for some honest $p_i$. Then, by the ignore rule,
there exists $p_j\in Cand_h^{(r+1)}$ such that
$support_i^{(r)}(j)\le k.$

Since $p_h$ is honest and $p_j\in Cand_h^{(r+1)}$, candidate $p_j$ was not removed by $p_h$
at the end of round $r$. Hence
$support_h^{(r)}(j)\ge 3k+1.$
Among these supporters, at least $2k+1$ are in $\hat\Pi^{4k+1}$, since
$\Pi_h^{4k+1}$ contains at most $k$ processes outside $\hat\Pi^{4k+1}$.

By Lemma~\ref{lem:pi hat intersect},
$|\Pi_i^{4k+1}\cap \hat\Pi^{4k+1}|\ge 3k+1$.
Thus, the $2k+1$ honest-prefix supporters seen by $p_h$ and the $3k+1$ honest-prefix processes
in $\Pi_i^{4k+1}$ intersect in at least
$(2k+1)+(3k+1)-(4k+1)=k+1$
processes. By the induction hypothesis, none of these honest processes is in $Ignore_i^{(r)}$.
Therefore
$support_i^{(r)}(j)\ge k+1$,
contradicting $support_i^{(r)}(j)\le k$.

Hence no honest process is added to any ignore set in round $r+1$, completing the induction.
\end{proof}

\begin{lemma}[Three-round elimination of the global honest minimum]
\label{lem:Cand_gettingSmaller_for_all_honest_formal-app}
Let $m^{(r)}= \min\left(\bigcup_{s\in C_{4k+1}\cap \mathcal H} Cand_s^{(r)}\right)$ for any $r\geq 1$ in which $\bigcup_{s\in C_{4k+1}\cap \mathcal H} Cand_s^{(r)}\neq\emptyset$.
Then $m^{(r)}\notin Cand_i^{(r+3)}$ for every honest core process $p_i\in C_{4k+1}$.
Consequently, $m^{(r)}<m^{(r+3)}$.
\end{lemma}

\begin{proof}
Write $m=m^{(r)}$, and fix an honest core process $p_i\in C_{4k+1} \cap H$.
If $m\notin Cand_i^{(r)}$, then since honest candidate sets only shrink,
$m\notin Cand_i^{(r+3)}$. So assume $m\in Cand_i^{(r)}$.

Consider any candidate $q<m$. By definition of $m$, no honest core process has $q$ in its
candidate set in round $r$. Therefore any process in
$\Pi_i^{4k+1}$ that reports $q$ in round $r$ must lie outside $\hat\Pi^{4k+1}$. There are at most $k$ such processes, hence $support_i^{(r)}(q)\le k$.
Thus, by the ignore-set update rule, any process that reports such a $q$ in round $r+1$ is added
to $Ignore_i^{(r+1)}$ before $support_i^{(r+1)}$ and $front_i^{(r+1)}$ are evaluated.
Consequently, no non-ignored process in round $r+1$ reports a candidate smaller than $m$. 
If $m\notin Cand_i^{(r+1)}$, we are done. Otherwise, $m$ is the smallest candidate appearing in
$p_i$'s active received candidate sets in round $r+1$. 
Now we will cover all cases and see that $m$ is added to the $RemoveSet$ on round $r+1$ or round $r+2$ and therefore $m\notin Cand_i^{(r+3)}$.
If $support_i^{(r+1)}(m)<3k+1$, then $m$ is part of the removed set of round $r+1$.
Otherwise, $support_i^{(r+1)}(m)\ge 3k+1$. Since $m$ is the smallest candidate appearing in
$p_i$'s active received candidate sets, whenever $m$ appears in a received set it is its minimum. Hence
$front_i^{(r+1)}(m)=support_i^{(r+1)}(m)\ge 3k+1$.
If $m=i$, then $front_i^{(r+1)}(i)\ge 3k+1$, and therefore $p_i$ sends a
$\textsf{LEADER}$ message in round $r+1$. Hence $i=m$ is included in
$UsedLeaders$ and is added to $RemoveSet$ in round $r+1$.
If $m\neq i$ and $p_i$ receives a $\textsf{LEADER}$ message from $p_m$ in round
$r+1$, then $m$ is included in $UsedLeaders$ and is added to $RemoveSet$ in
round $r+1$. Therefore, if $m$ survives round $r+1$, it must be that either
$m\neq i$ and $p_i$ did not receive a $\textsf{LEADER}$ message from $p_m$ in
that round. In round $r+2$, one of the following happens:
\begin{itemize}
    \item either $support_i^{(r+2)}(m)<3k+1$, in which case $m$ is added to $RemoveSet$;
    \item or $support_i^{(r+2)}(m)\ge 3k+1$. Since $m$ was the smallest identifier appearing in the active received candidate sets in round $r+1$, every $q<m$ had $support_i^{(r+1)}(q)=0\le k$. Hence any process that reports such a $q$ in round $r+2$ is added to $Ignore_i^{(r+2)}$ before $front_i^{(r+2)}$ and $support_i^{(r+2)}$ are evaluated. Therefore, among the non-ignored processes in round $r+2$, $m$ is again the minimum of every received set containing it, so $front_i^{(r+2)}(m)\ge 3k+1$.
    Since also $front_i^{(r+1)}(m)\ge 3k+1$, candidate $m$ is added to $RemoveSet$ by the consecutive-front cleanup rule. 
\end{itemize}
Thus in all cases $m$ will be removed in round $r,r+1$ or $r+2$ so $m\notin Cand_i^{(r+3)}$.
Since $p_i$ was arbitrary, $m$ is absent from every honest core candidate set in round $r+3$. Consequently, if $m^{(r+3)}$ is defined, then $m^{(r)}<m^{(r+3)}$.
\end{proof}

\begin{lemma}\label{lem:remain-candidate-app}
    Assume that some process $p^\star$ is initially in the candidate set of every honest core process and that there is no round $r$ such that there is some honest core process $p_h$ such that $front_h^{(r)}(p^{\star}) \geq 3k+1$. Then $p^\star$ stays in the candidate sets of all honest core processes forever.
\end{lemma}

\begin{proof}
We prove this claim by induction. Initially, $p^\star$ is in all candidate sets by construction. 
Now consider any round $r$ in which $p^\star$ belongs to every honest core
candidate set at the beginning of the round. Let $p_i$ be an honest core process.
By Lemma~\ref{lem:pi hat intersect}, $p_i$ receives candidate sets from
at least $3k+1$ honest core processes in $\hat\Pi^{4k+1}$, and by
Lemma~\ref{lem:ignore_set_safety} none of these processes is ignored by $p_i$.
Since all of them still contain $p^\star$, we have $support_i^{(r)}(p^\star)\ge 3k+1$.
Thus $p^\star$ is not added to the \texttt{RemoveSet} in the first set on line~\ref{line:cleanup}. It is also not added in the second set (i.e., the \texttt{UsedLeaders} set), because $p^\star$ is honest, and by our assumption, no honest
process ever sees
$front^{(r)}(p^\star)\ge 3k+1$, so $p^\star$ does not send a
\textsf{LEADER} message. Finally, it is not added to the last set that makes up the \texttt{RemoveSet}, since that rule requires
$p^\star$ to be the front runner at least $3k+1$ in some round, which does not happen, again by our assumption. Hence $p^\star$ remains
in every honest core candidate set in round $r+1$, proving the claim.
\end{proof}

\begin{lemma}\label{lem:eventually-good-front-app}
    There exists a round $r$ and an honest core process $p_h$ such that
\[
front_h^{(r)}(p^\star)\ge 3k+1,
\]
where $p^\star$ is the minimum-ID honest core process that is classified positively by every honest core process.
\end{lemma}
\begin{proof}
We prove the claim by contradiction. Suppose the claim does not hold. Note that by construction, a core process classified positively by an honest core process $p_i$ appears in its candidate set in round 1. By Lemma~\ref{lem:remain-candidate-app}, $p^\star$ remains in all honest core candidate sets forever.

Therefore the global honest candidate set is never empty. By
Lemma~\ref{lem:Cand_gettingSmaller_for_all_honest}, the global honest minimum
strictly increases every constant number of rounds. Since, by
Lemma~\ref{lem:pk-blocking-k}, fewer than $8k$ candidates smaller than
$p^\star$ ever appear in honest core candidate sets, eventually no honest core
candidate set contains a candidate smaller than $p^\star$. At that point,
$p^\star$ is the minimum candidate in every honest core candidate set. Furthermore, recall that by Lemma~\ref{lem:ignore_set_safety}, no honest process is ignored by any honest process. Hence,
every honest core process sees $p^\star$ as the front-runner in at least
$3k+1$ non-ignored honest core candidate sets, contradicting our assumption.
Thus such a round exists.
\end{proof}

\begin{lemma}
\label{lem:goodRoundExists-app}
 Let $p^\star$ be the minimum-ID honest core process that is classified positively by every honest core process. Then either some honest core process decides before $p^\star$ acts as a leader, or there exists a round in which $p^\star$ establishes a good synchronizing round.
\end{lemma}

\begin{proof}
We prove that, unless an honest core process decides earlier, the process $p^\star$ eventually becomes an acceptable leader for all honest core processes.

Let $r$ be the first round in which some honest core process $p_h$ has $front_h^{(r)}(p^\star)\ge 3k+1$. By Lemma~\ref{lem:eventually-good-front-app}, such a round exists.

Let $\mathit{GoodVoters_h}$ be the subset of processes in $\hat\Pi^{4k+1}$ listened to by $p_h$ that have $p^\star$ as their front runner. Formally, 
\[
GoodVoters_h=\{\,x\in(\Pi_h^{4k+1}\setminus Ignore_h^{(r)})\cap\hat\Pi^{4k+1}
: p^\star=\min Cand_x^{(r)}\,\}.
\]
Note that since, by Lemma~\ref{lem:pi hat intersect}, at most $k$ processes in $p_h$'s listening set lie outside $\hat\Pi^{4k+1}$,
$|GoodVoters_h|\ge 2k+1$.

By Lemma~\ref{lem:pi hat intersect}, for any honest process $p_i$, $|\Pi_i^{4k+1}\cap \hat\Pi^{4k+1}|\ge 3k+1$. Since both $GoodVoters_h$ (of size at least $2k+1$) and $\Pi_i^{4k+1}\cap \hat\Pi^{4k+1}$ are subsets of $\hat\Pi^{4k+1}$ (of size $4k+1$), inclusion--exclusion yields
\[
|\Pi_i^{4k+1}\cap GoodVoters_h\cap \mathcal{H}| = |\Pi_i^{4k+1}\cap GoodVoters_h| \ge (2k+1)+(3k+1)-(4k+1)=k+1.
\]
By Lemma~\ref{lem:ignore_set_safety-app}, these honest processes are not ignored by $p_i$, and therefore $front_i^{(r)}(p^\star) \geq k+1$.
Furthermore, by the choice of $r$ and Lemma~\ref{lem:remain-candidate-app}, $p^\star$ is still in every candidate set in round $r$, and thus $support_i^{(r)} \geq 3k+1$.  

We distinguish two cases.

\medskip
\noindent\textbf{Case 1:}
\[
front_{p^\star}^{(r)}(p^\star)\ge 3k+1.
\]
Then $p^\star$ sends $\textsf{LEADER}(v_{p^\star}^{tmp})$ in round $r$. Since $\textsf{LEADER}$ messages are sent to all processes, every honest process receives it. As established above, for every honest $p_i$,
\[
p^\star\in Cand_i^{(r)},\qquad
\mathrm{support}_i^{(r)}(p^\star)\ge 3k+1,\qquad
\mathrm{front}_i^{(r)}(p^\star)\ge k+1.
\]
Thus every honest process with grade $0$ accepts the value sent by $p^\star$ in round $r$. Further, by Lemma~\ref{lem:king value consistency}, all processes with $maxg = 1$ have the same $newval$ in this round. So $r$ is a good synchronizing round, as desired.

\medskip
\noindent\textbf{Case 2:}
\[
front_{p^\star}^{(r)}(p^\star)<3k+1.
\]
We claim that every candidate $q<p^\star$ is removed from every honest candidate set at the end of round $r$.

Fix an honest process $p_i$ and a candidate $q<p^\star$. Suppose, towards contradiction, that
\[
\mathrm{support}_i^{(r)}(q)\ge 3k+1.
\]
Then at least $2k+1$ counted supporters of $q$ lie in $\hat\Pi^{4k+1}$. Let
\[
Q_q=\{\,x\in(\Pi_i^{4k+1}\setminus Ignore_i^{(r)})\cap\hat\Pi^{4k+1}
: q\in Cand_x^{(r)}\,\}.
\]
Then $|Q_q|\ge 2k+1$. Since $GoodVoters_h,Q_q\subseteq\hat\Pi^{4k+1}$,
\[
|GoodVoters_h\cap Q_q|\ge (2k+1)+(2k+1)-(4k+1)=1.
\]
Thus some honest process $p_s$ has both $p^\star=\min Cand_s^{(r)}$ and $q\in Cand_s^{(r)}$, impossible because $q<p^\star$. Hence
\[
support_i^{(r)}(q)<3k+1,
\]
so every honest process removes every such $q$ in line~\ref{line:cleanup} in the first subset of \texttt{RemoveSet}.

Process $p^\star$ itself is not removed at the end of round $r$, since it is supported by all honest processes, did not act as leader in this round, and has been the front runner of at least $3k+1$ candidate sets in any honest core process's listening sets for at most 1 round, by the choice of $r$. 

Therefore, at the beginning of round $r+1$, every honest candidate set contains
$p^\star$ and no honest candidate set contains a smaller candidate. Since honest
processes are never ignored by Lemma~\ref{lem:ignore_set_safety}, every honest
process sees at least $3k+1$ non-ignored honest sets whose minimum is $p^\star$.
In particular,
\[
front_{p^\star}^{(r+1)}(p^\star)\ge 3k+1,
\]
so $p^\star$ sends $\textsf{LEADER}(newval)$ in round $r+1$. The same
acceptance argument as in Case 1 shows that every honest process with grade $0$
accepts it. Further, by Lemma~\ref{lem:king value consistency}, all processes with $maxg = 1$ have the same $newval$ in this round. Therefore round $r+1$ is a good synchronizing round, as desired.
\end{proof}

\begin{lemma} 
\label{lem:goodRoundCauseTermination}
If no honest core process decides before round $r$, and round $r$ is a good synchronizing round, then all honest core processes decide in round $r+1$, and all decide the same value. 
\end{lemma} 
\begin{proof}
Suppose no honest core process decides before round $r$, and round $r$ is a good synchronizing round with $p^\star$ acting as leader, sending $newval$ in its proposal. By definition of a good synchronizing round, after the grade-update step of round $r$, every honest core process with $maxg=1$ already has $newval$ as its current value, while every honest core process with $maxg=0$ accepts the value sent by $p^\star$ in the leader step. Hence, after the leader step and value update of round $r$, all honest core processes start round $r+1$ with the same value. By the same validity argument used in Lemma~\ref{lem:DecideInOneMoreRound}, the first graded agreement invocation in round $r+1$ returns this value with grade $1$ to every honest core process, and the second invocation on unanimous input $1$ returns $(1,1)$. Hence all honest core processes decide in round $r+1$, and all decide the same value.
\end{proof}

\begin{theorem}[Correctness]
\label{thm:pk-correctness-app}
The Predictive Core-Leader Agreement (PCLA) protocol satisfies validity, agreement, and termination.
\end{theorem}

\begin{proof}
\textbf{Validity.}
Suppose all honest processes start with the same input value $v$. Then in particular all honest core processes in $C_{4k+1}$ start with value $v$. In the first round, the first invocation of graded agreement is executed on unanimous honest core input $v$, and therefore every honest core process obtains $(v,1)$. The second invocation is then executed on unanimous input $1$, hence every honest core process obtains $(1,1)$, broadcasts $\textsf{DECIDE}(v)$, and outputs $v$. By Lemma~\ref{lem:pk-dissemination}, every honest process outside $C_{4k+1}$ eventually decides $v$ as well. Therefore validity holds for all honest processes.

\medskip
\noindent\textbf{Agreement.}
We distinguish two cases.

\smallskip
\noindent\emph{Case 1: Some honest core process outputs before any good synchronizing round occurs.}

Let $r$ be the first round in which an honest core process outputs. By Lemma~\ref{lem:DecideInOneMoreRound}, every honest core process that outputs in round $r$ outputs the same value, and every honest core process that does not output in round $r$ enters round $r+1$ with that same value and with grade $1$. Consequently, all remaining honest core processes output in round $r+1$ with the same value. Since outputting core processes broadcast their decision values, Lemma~\ref{lem:pk-dissemination} implies that every honest non-core process eventually decides that same value. Therefore agreement holds for all honest processes.

\smallskip
\noindent\emph{Case 2: No honest core process outputs before any good synchronizing round occurs.}

By Lemma~\ref{lem:goodRoundExists-app}, a good synchronizing round exists; let $r$ be the first such round. Since no honest core process outputs before round $r$, Lemma~\ref{lem:goodRoundCauseTermination} implies that all honest core processes output in round $r+1$, and all output the same value. They all broadcast this value, and by Lemma~\ref{lem:pk-dissemination}, every honest process outside $C_{4k+1}$ eventually decides the same value. Therefore agreement holds for all honest processes.

\medskip
\noindent\textbf{Termination.}
Again, we distinguish two cases.

If some honest core process outputs before any good synchronizing round occurs, then by Lemma~\ref{lem:DecideInOneMoreRound}, every remaining honest core process outputs in the next round. By the protocol, outputting core processes continue for one additional iteration and then terminate. Once all honest core processes have output the same value, they have all broadcast $\textsf{DECIDE}$ for that value, so Lemma~\ref{lem:pk-dissemination} implies that every honest non-core process eventually decides. Hence every honest process eventually terminates.

Otherwise, no honest core process outputs before any good synchronizing round occurs. By Lemma~\ref{lem:goodRoundExists-app}, a good synchronizing round exists; let $r$ be the first such round. Lemma~\ref{lem:goodRoundCauseTermination} implies that all honest core processes output in round $r+1$. By the protocol, they continue for one additional iteration and then terminate. Once they broadcast their common decision value, Lemma~\ref{lem:pk-dissemination} implies that every honest process outside $C_{4k+1}$ also eventually decides. Hence every honest process eventually terminates.
\end{proof}

\section{Lower bounds}
\label{app:lowerbound}
\begin{theorem}
Assume $t \ge k \ge 1$. 
For every deterministic synchronous algorithm that solves Byzantine broadcast with up to $t$ Byzantine faults in the local $k$-error prediction model, there exists an all-honest execution in which all processes receive the same perfectly accurate prediction vector, and in which honest processes send $\Omega(nk)$ messages.
\end{theorem}
\begin{proof}
Let $P$ be a deterministic Byzantine broadcast protocol in the local $k$-error prediction model. 
Without loss of generality, let $p_1$ be the designated sender. Consider the following two executions:
\begin{itemize}
    \item $E_0$: all processes are honest, every process receives the prediction vector $(1,\dots,1)$, and the sender's input is $0$;
    \item $E_1$: all processes are honest, every process receives the prediction vector $(1,\dots,1)$, and the sender's input is $1$.
\end{itemize}
In both executions, the common prediction vector is perfectly accurate.

Assume, toward a contradiction, that in both $E_0$ and $E_1$ the honest processes send fewer than $nk/8$ messages. 
For every process $p_j \neq p_1$, let $\mathrm{touch}(p_j)$ be the total number of messages incident to $p_j$ over both executions $E_0$ and $E_1$ together, that is, messages either sent by $p_j$ or received by $p_j$ in either one of the executions.
Since each message contributes at most $2$ to the total touch-count, we have
\[
\sum_{j \neq 1} \mathrm{touch}(p_j)
< 2\cdot \frac{nk}{8} + 2\cdot \frac{nk}{8}
= \frac{nk}{2}.
\] 
Therefore, by averaging over the $n-1$ processes distinct from the sender, there exists a process $p_i \neq p_1$ such that
\[
\mathrm{touch}(p_i)
< \frac{nk}{2(n-1)}
< k,
\]
where the last inequality holds for $n \ge 3$.
Let $\Lambda$ be the set of all processes $p_j\neq p_i$ that either send a message to $p_i$ or receive a message from $p_i$ in at least one of the two executions $E_0,E_1$. By definition,
$|\Lambda| \le \mathrm{touch}(p_i) < k$. Since $t \ge k$, it is valid to choose all processes in $\Lambda$ as Byzantine.
Now define a third execution $E_{\mathsf{bad}}$ as follows:
\begin{itemize}
    \item every process in $\Lambda$ is Byzantine;
    \item every process outside $\Lambda$ is honest, including $p_i$;
    \item if the designated sender $p_1\notin\Lambda$, then it is honest with input $0$; if $p_1\in\Lambda$, then it is Byzantine;
    \item every honest process receives the prediction vector $(1,\dots,1)$.
\end{itemize}
This execution satisfies the local $k$-error prediction assumption: the faulty set is exactly $\Lambda$, every honest process receives the all-ones prediction vector, and therefore each honest prediction vector has exactly $|\Lambda|\le k$ false positives and no false negatives.

The Byzantine processes in $\Lambda$ behave as in execution $E_0$ toward every honest process other than $p_i$, and behave as in execution $E_1$ toward $p_i$. If $p_1\notin\Lambda$, then the honest sender has input $0$, as in $E_0$. This is consistent because, by construction, every process that sends a message to $p_i$ or receives a message from $p_i$ in either $E_0$ or $E_1$ belongs to $\Lambda$.
Consequently:
\begin{itemize}
    \item for every honest process $p_j \neq p_i$, the local view of $p_j$ in $E_{\mathsf{bad}}$ is indistinguishable from its view in $E_0$, where the honest sender has input $0$; hence $p_j$ outputs $0$;
    \item the local view of $p_i$ in $E_{\mathsf{bad}}$ is indistinguishable from its view in $E_1$, where the honest sender has input $1$; hence $p_i$ outputs $1$.
\end{itemize}

This violates agreement in $E_{\mathsf{bad}}$, since both $p_i$ and the honest processes outside $\Lambda$ are honest. Therefore, at least one of $E_0$ and $E_1$ must contain at least $nk/8$ honest messages. Hence there exists an all-honest execution with a common perfectly accurate prediction vector in which honest processes send $\Omega(nk)$ messages.
\end{proof}

\subsection{Reductions from Classical Byzantine Broadcast}

\begin{theorem}[Message Lower Bound Reduction]
\label{prop:msg-reduction}
Any deterministic message lower bound for synchronous Byzantine broadcast with up to $t$ Byzantine faults also applies to Byzantine broadcast in the local $k$-error prediction model if $t\le k$.
\end{theorem}
\begin{proof}
Consider any execution with at most $t$ Byzantine faults, and give every honest process the all-ones prediction vector $a_i = (1,\dots,1)$. Since there are at most $t \le k$ faulty processes, each honest process makes at most $k$ prediction errors, all of them false positives. Thus, the execution is valid in the local $k$-error prediction model. Therefore, on this family of executions, any protocol for the local $k$-error prediction model must also solve ordinary synchronous Byzantine broadcast with up to $t$ Byzantine faults, inheriting all deterministic message lower bounds.
\end{proof}

\begin{theorem}[Round Lower Bound Reduction]
\label{prop:round-reduction}
Let $\ell \le \min\{t,k\}$. Any deterministic round lower bound for synchronous Byzantine broadcast with up to $\ell$ Byzantine faults also applies to Byzantine broadcast in the local $k$-error prediction model. In particular, the worst-case round complexity is at least $\min\{t,k\}+1$.
\end{theorem}
\begin{proof}
The proof follows by the same reduction as in Theorem~\ref{prop:msg-reduction}, restricting the fault count to $\ell \le \min\{t,k\}$.
\end{proof}

\end{document}